\documentclass[aps, pra, twocolumn,superscriptaddress,groupedaddress]{revtex4}

\usepackage{array}
\usepackage{graphicx}
\usepackage{dcolumn}
\usepackage{bm}
\usepackage{amsmath}
\usepackage{amsfonts}
\usepackage{amssymb}
\usepackage{amstext}   
\usepackage{amsthm}
\usepackage{dsfont}
\usepackage{color}
\usepackage{float}
\usepackage[utf8]{inputenc}
\usepackage[outdir=./]{epstopdf}
\usepackage{nameref}
\usepackage{subfloat}
\usepackage[colorlinks = true, citecolor = blue]{hyperref}
\usepackage[figure]{hypcap}
\usepackage[colorinlistoftodos]{todonotes}
\usepackage[writekey = false]{notes2bib}
\usepackage{mathtools}

\newcommand{\ket}[1]{\ensuremath{\left|#1\right\rangle}}

\newcommand{\dg}{\dagger}
\newcommand{\va}{\vec{\alpha}}
\newcommand{\vb}{\vec{\beta}}
\newcommand{\vg}{\vec{\gamma}}
\newcommand{\ve}{\vec{\eta}}
\newcommand{\vl}{\vec{\lambda}}
\newcommand{\vS}{\vec{S}}
\newcommand{\p}{\prime}

\DeclareMathOperator{\tr}{tr}

\begin{document}

\title{Many-body-localization transition in a universal quantum circuit model}

\author{Adrian Chapman}
\email{akchapman@unm.edu}
\author{Akimasa Miyake}
\email{amiyake@unm.edu}
\affiliation{Center for Quantum Information and Control, Department of Physics and Astronomy, University of New Mexico, Albuquerque, NM 87131, USA} 

\date{\today}

\begin{abstract}

The dynamical generation of complex correlations in quantum many-body systems is of renewed interest in the context of quantum chaos, where the out-of-time-ordered (OTO) correlation function appears as a convenient measure of scrambling. To detect the the transition from scrambling to many-body localization, the latter of which has limited dynamical complexity and is often classically simulatable, we develop both exact and approximate methods to compute OTO correlators for arbitrary universal quantum circuits. We take advantage of the mapping of quantum circuits to the dynamics of interacting fermions in one dimension, as Gaussian time evolution supplemented by quartic interaction gates. In this framework, the OTO correlator can be calculated exactly as a superposition of exponentially many Gaussian-fermionic trajectories in the number of interaction gates. We develop a variationally-optimized, Gaussian approximation to the spatial propagation of an initially-local operator by restriction to the fastest-traveling fermionic modes, in a similar spirit as light-front computational methods in quantum field theory. We demonstrate that our method can detect the many-body localization transitions of generally time-dependent dynamics without the need for perturbatively weak interactions. \\

\end{abstract}

\maketitle

\section{Introduction}

By now it is well-understood that quantum effects play a prominent role for information propagation in many-body systems. Namely, the rate at which local disturbances propagate into nonlocal degrees of freedom --- or scramble --- under unitary dynamics is limited by the Lieb-Robinson bound \cite{lieb1972finite}. This endows the system with an effective ``speed of light," even without any invocation of relativity \emph{a priori}. This uniquely quantum phenomenon follows from the locality structure of the Hamiltonian alone, and therefore is a ubiquitous property among quantum lattice systems.

A natural question, then, is how Lieb-Robinson-bounded propagation of quantum information will affect the performance of a quantum computer. As any practical realization of a quantum circuit will naturally possess some inherent notion of locality due to its connectivity structure, it seems obvious that there is a minimum circuit depth before the system will be able to access any given extensively nonlocal degree of freedom. This is simply the number of gate layers needed for the support of a local observable to interact with every qubit in the system. However, it may be possible for a more stringent bound to hold due to the particular nature of the dynamics as well. An analogous situation can be seen in the \emph{many-body-localized} regime for Hamiltonian dynamics in the presence of a disordered local field and perturbatively weak interactions \cite{gornyi2005interacting, basko2006metal, imbrie2016mbl, serbyn2013local}. In such systems, the support of a disturbance will propagate logarithmically, rather than linearly, with time \cite{huse2014phenomenology, chowdhury2016slow, deng2017logarithmic, nanduri2014entanglement}. The minimum time needed for the system to access extensively nonlocal degrees of freedom in this case is therefore exponential in the system size. Since strong quantum correlations cannot be built quickly, such systems admit many properties which are classically simulatable \cite{bravyi2006liebrobinson, znidaric2008manybody, bardarson2012unbounded, kim2014local, cirac2017dynamics}. We therefore ask whether a transition to many-body-localized behavior exists in quantum circuits. Such a transition would be tantamount to a \emph{complexity transition}, for which a full understanding would be of great importance. Furthermore, the dynamics of quantum circuits is closely related to that of periodically-driven Floquet systems \cite{chandran2015semiclassical, keser2016dynamical}, where it has been shown that many-body-localized behavior indeed survives \cite{abanin2016theory, ponte2015mbl, ponte2015periodically, mierzejewski2017mbl}.

A recent tool developed for the purpose of accessing many-body-scrambling is the out-of-time-ordered (OTO) correlator, which was introduced by Kitaev to model the fast-scrambling behavior of black holes \cite{kitaev2015kitp}. Since then, the OTO correlator has enjoyed success in describing the scrambling behavior of chaotic quantum systems. It has been used, for example, to study chaotic behavior in random quantum circuit models \cite{gullans2018entanglement, zhou2018emergent,  zhou2018operator, jonay2018coarsegrained, xu2018locality, sunderhauf2018localisation, nahum2018operator, keyserlingk2017operator} --- including those with conservation laws \cite{rakovszky2018diffusive, khemani2017operator} --- and the related dynamics of random-matrix models \cite{gharibyan2018onset, kos2018manybody, chen2018operator}. Conversely, it has been shown that the OTO correlator is effective at detecting the \emph{absence of scrambling}, as seen in the many-body localized phase \cite{fan2017oto, he2016characterizing, chen2016universal, xiao2016oto, slagle2017oto}. In fact, it is argued in Ref.~\cite{huang2016oto} that the OTO correlator is uniquely-suited to this task. Such properties make the OTO correlator an ideal diagnostic for the many-body-localization transition in quantum circuits and ensembles thereof. Nevertheless, utilizing this quantity to detect localization without \emph{a priori} knowledge of such behavior in the general, single-shot regime remains a challenge, since it would in principle require full simulation over an exponentially large Hilbert space. In Ref.s~\cite{xu2018accessing, sahu2018scrambling}, the authors utilize matrix product operators, truncated to low bond dimension, to approximate the Heisenberg operator time evolution and calculate the OTO correlator for scrambling and localizing systems. This method can be viewed as a generalization of performing gate cancellations outside of the trivial lightcone of a quantum circuit by taking the particular circuit dynamics into account, and approximating the circuit inside the ``true lightcone" by one of low depth. In Ref.~\cite{sunderhauf2018localisation}, the authors observe a many-body localization transition in a Floquet model with Haar random local unitaries together with disordered 2-qubit interactions, for which they employ a similarly clever tensor network contraction scheme to reduce the complexity of their quantity (which is not the OTO correlator) by an exponential factor, though it is still exponential overall. They also demonstrate localizing behavior in a Floquet circuit model of random Gaussian-fermionic circuits, which admits an efficient classical simulation.

In this paper, we take advantage of the fact that any time evolution can be written in terms of dynamics of interacting fermions \cite{bravyi2002fermionic}, so that the OTO correlator may be computed as a determinental formula as studied in our previous work for non-interacting fermions \cite{chapman2017classical}. We first derive an exact formula for the OTO correlator for universal quantum circuits, expressed in terms of Gaussian-fermionic evolution together with fermionic ``interaction" gates, as a superposition of exponentially many free-fermion trajectories. This formula is an alternating series of determinants of sub-matrices of an orthogonal, symmetric matrix, which reflects the fact that our fermionic interaction gates only permit transitions between certain configurations of fermions. In a similar spirit as light-front computational methods in quantum field theory, we restrict our formula to keep track of only the fastest traveling modes, allowing us to replicate the action of an interaction gate by that of a Gaussian-fermionic circuit coupling to a set of ancillary modes and approximate the time-evolution efficiently (i.e. in-terms of a single determinant). We apply our algorithm to a universal quantum circuit model consisting of alternating layers of non-interacting fermion evolution, and interaction gates coupling alternating subsets of qubits, where we observe a transition to many-body-localized behavior as we increase the disorder strength. Though we consider an ensemble-averaged Floquet model for ease of presentation in this work, we emphasize that neither of these is necessary for our algorithm. Our algorithm can be applied for any one-dimensional nearest-neighbor quantum circuit, without need to work in the perturbatively-interacting regime, and without the need for super-computing resources.

\section{Background}
\label{sec:bkgrnd}

\subsection{A. Out-of-Time-Ordered Correlator}
\label{sec:otoca}

Our figure of interest is the infinite-temperature out-of-time-ordered (OTO) correlator, defined between two observables $A$ and $B(t) \equiv U B U^{\dg}$ for a system of Hilbert-space dimension $d$ and unitary time evolution $U$ as

\begin{align}
\mathcal{C}_{AB}(t) \equiv (4d)^{-1/2}||[A, B(t)]||_F \mathrm{,}
\label{eq:OTOCdef}
\end{align}

\noindent where $||A||_F \equiv \sqrt{\tr(A^{\dg} A)}$ is the Frobenius norm. When $A$ and $B$ are Hermitian and unitary operators (e.g. qubit Pauli observables), we have the relation 

\begin{align}
\mathcal{C}_{AB}(t)^2 = \frac{1}{2}\left\{1 - d^{-1} \tr\left[AB(t)AB(t)\right]\right\}. 
\label{eq:cfrelation}
\end{align}

\noindent Here, we will choose $A = X_{\lfloor n/2 \rfloor}$ and $B = Z_{s}$ for qubit $s \in \{1, 2, \dots, n\}$. ``Infinite temperature" refers to the fact that the trace in Eq.~(\ref{eq:cfrelation}) is the trace inner product between $A B(t) A B(t)$ and the infinite-temperature Gibbs state $d^{-1} I$. This trace term is sometimes referred to as the OTO correlator in the literature, and $\mathcal{C}_{AB}$ is called the OTO \emph{commutator}. Here we will refer to either quantity as the OTO correlator, as our meaning will be clear from context. The normalization in Eq.~(\ref{eq:OTOCdef}) is such that $\mathcal{C}_{AB} \in [0, 1]$. $\mathcal{C}_{AB}$ is further bounded by the conventional Lieb-Robinson commutator norm by the operator-norm inequality $||A||_F \leq \sqrt{d}||A||_2$, as

\begin{align}
\mathcal{C}_{AB}(t) \leq \frac{1}{2} ||[A, B(t)]||_2 \lesssim ||A||_2 ||B||_2 e^{-\eta \left(d_{AB} - v t \right)} \mathrm{,}
\label{eq:LRbound}
\end{align}

\noindent where $d_{AB}$ is the initial lattice distance between $A$ and $B$ ($d_{AB} = |\lfloor n/2 \rfloor - s|$ for our choices of $A$ and $B$), and the second inequality is the Lieb-Robinson bound \cite{lieb1972finite}. This bound confines the support of $B(t)$ to within an effective ``lightcone" of speed $v$, outside of which the amplitudes of $B(t)$ in a local operator basis decay exponentially.

As stated above, a more stringent bound than Eq.~(\ref{eq:LRbound}) holds for many-body-localized systems. Namely, support in such systems is confined to within a \emph{logarithmic lightcone}

\begin{align}
\mathcal{C}_{AB}(t) \lesssim ||A||_2 ||B||_2 e^{-\eta \left(d_{AB} - v \ln{t} \right)} \mathrm{.}
\label{eq:loglightcone}
\end{align}

\noindent That is, disturbances take exponential time to propagate a given distance. This behavior is intimately related to a logarithmic spreading of entanglement \cite{fan2017oto}, which is a signature of many-body localization \cite{slagle2017oto, chowdhury2016slow, kim2014local, bardarson2012unbounded, znidaric2008manybody, serbyn2013local, cirac2017dynamics, deng2017logarithmic, imbrie2016mbl} (further, it has been shown to be distinct from the Anderson-localized phase \cite{he2016characterizing, chen2016universal}, in which the lightcone width is constant in time \cite{anderson1958absence}). 

Though Eq.~(\ref{eq:LRbound}) is an upper bound, we expect $\mathcal{C}_{AB}(t)$ to give a good heuristic for the lightcone, and in fact it was shown in Ref.~\cite{huang2016oto} that the OTO correlator can detect the logarithmic lightcone where more conventional, two-point correlators cannot. Averages of the OTO correlator are also useful, since they are be related to the more familiar second R\'{e}nyi entanglement entropy

\begin{align}
S_M^{(2)} \equiv -\log{\tr_{M}{\left(\tr_{\overline{M}}\rho\right)^2}}\mathrm{,} 
\end{align}

\noindent with respect to a subsystem $M$ for $\rho$ an infinite-temperature Gibbs state quenched to the eigenbasis of an operator $A$. This relation is the so-called ``OTOC-RE" theorem \cite{fan2017oto} 

\begin{align}
\exp{\left[-S_M^{(2)}(t)\right]} = \sum_{B \in \overline{M}} \tr{\left[B(t) \Pi_A B(t) \Pi_A \right]}\mathrm{,}
\end{align}

\noindent where $\Pi_A$ is the trace-normalized projector onto the eigenbasis of $A$, the sum is taken over a local operator basis on $\overline{M}$, and normalization is chosen such that $\sum_{B \in \overline{M}} B_{ij} B_{lm} = \delta_{im} \delta_{lj}$, $\tr{\Pi_A} = 1$. This connection was extended to operator entanglement in Ref.~\cite{xu2018accessing}, building off of the work in \cite{hosur2016chaos}, through the bound

\begin{align}
S^{(2)}_{M}(t) \leq -\log \left(1 - \frac{1}{2} \sum_{\{P_j | j \in M\}} \mathcal{C}^2_{A P_j}(t)\right) \mathrm{,}
\label{eq:S2operatorbound}
\end{align}

\noindent where here, the operator entanglement is that of the state related to the original operator by contracting the operator on one side, say, subsystem $r$ of the infinite-temperature $n$-qubit thermofield double state

\begin{align}
\ket{\Phi} = 2^{-n/2} \sum_{j \in \{0, 1\}^{\times n}} \ket{j}_r \otimes \ket{j}_{\tilde{r}}
\label{eq:thermofielddoubledef}
\end{align}

\noindent and the sum taken in Eq.~(\ref{eq:S2operatorbound}) is over all single-qubit Pauli operators $\{I, X, Y, Z\}$ supported on $M$. Though a bounded second R\'{e}nyi entropy does not guarantee a low-bond-dimension matrix product operator approximation, it is argued in Ref.s~\cite{xu2018accessing, sahu2018scrambling} that the lightcone envelope can be well-approximated using such a technique, since any disturbance due to truncating the matrix product operator to limited bond dimension cannot itself propagate outside of the lightcone. See also Ref.s~\cite{hosur2016chaos, cotler2017chaos}, where the averaged OTO correlator has been related to the tripartite second R\'{e}nyi mutual information and the spectral form factor, respectively.

\subsection{B. Gaussian Fermionic Evolution}
\label{sec:gfeb}

We next give a brief review of Gaussian fermionic evolution (also known as matchgate circuits, see Ref.s~\cite{knill2001fermionic, terhal2002classical, bravyi2002fermionic, bravyi2006universal, jozsa2008matchgates, melo2013power, brod2014computational} for further details), which we define in terms of the Jordan-Wigner transformation from Pauli observables on $n$ qubits to \emph{Majorana operators} on $2n$ fermionic modes, as

\begin{align}
c_{2j - 1} = Z^{\otimes (j - 1)} X_j \mbox{\hspace{7mm}} c_{2j} = Z^{\otimes (j - 1)} Y_j \mathrm{.}
\label{eq:JWtransformation}
\end{align}

\noindent These operators satisfy the canonical anticommutation relations $\{c_{\mu}, c_{\nu}\} = 2\delta_{\mu, \nu} I$, where $\delta_{\mu, \nu}$ is the Kronecker delta. Gaussian fermionic unitaries are those of the form $U_{g} = \exp{\left(\mathbf{c}^{\mathrm{T}} \cdot \mathbf{h} \cdot \mathbf{c} \right)}$, where $\mathbf{c}$ is the column vector of Majorana operators. Unitarity and the canonical anticommutation relations restrict $\mathbf{h}$ to be a real, antisymmetric matrix without loss of generality.

Majorana operators are preserved under commutation with quadratic terms, as 

\begin{align}
[\mathbf{c}^{\mathrm{T}} \cdot \mathbf{h} \cdot \mathbf{c}, c_{\mu}] = (-4\mathbf{h} \cdot \mathbf{c})_{\mu} \mathrm{.}
\label{eq:covariantommutator}
\end{align}

\noindent This implies that, under Gaussian evolution, we have

\begin{align}
U_g^{\dg} c_{\mu} U_{g} = \left(e^{-4 \mathbf{h}} \cdot \mathbf{c} \right)_{\mu} \mathrm{.}
\end{align}

\noindent As $\mathbf{h}$ is an antisymmetric matrix, $\mathbf{u} \equiv e^{-4 \mathbf{h}} \in \mathrm{SO}(2n)$, and so Majorana operators form a representation of the group $\mathrm{SO}(2n)$ under Gaussian fermionic evolution. Additional representations can be constructed from ordered products of the Majorana operators --- called \emph{Majorana configuration operators} --- which we define as

\begin{align}
C_{\va} \equiv c_{\alpha_1} c_{\alpha_2} \dots c_{\alpha_k} \rm{,}
\end{align}

\noindent where $\vec{\alpha}$ is an \emph{ordered k-tuple}, for which $1 \leq \alpha_1 < \alpha_2 < \dots < \alpha_{k} \leq 2n$. Under Gaussian fermionic evolution, the Majorana configuration operators transform as

\begin{align}
U_{g}^{\dg} C_{\va} U_{g} = \sum_{\vb} \det\left(\mathbf{u}_{\va \vb}\right) C_{\vb} \mathrm{,}
\label{eq:gfevolution}
\end{align}

\noindent where $\mathbf{u}_{\vec{\alpha} \vec{\beta}}$ is the submatrix of $\mathbf{u}$ given by taking the rows indexed by $\va$ and the columns indexed by $\vb$. Matrices of amplitudes $\{\det\left(\mathbf{u}_{\va \vb}\right)\}_{\va \vb}$, whose elements are indexed by $k$-tuples, form a homomorphism of $\mathrm{SO}(2n)$ by the \emph{Cauchy-Binet formula}

\begin{align}
\sum_{\vb} \det\left[\left(\mathbf{u}_1\right)_{\va \vb}\right]\det\left[\left(\mathbf{u}_2\right)_{\vb \vg}\right] = \det\left[\left(\mathbf{u}_1\mathbf{u}_2\right)_{\va \vg}\right] \mathrm{.}
\label{eq:CBformula}
\end{align}

\noindent Eq.~(\ref{eq:CBformula}) will prove useful for calculating operator amplitudes for arbitrary Pauli operators under Gaussian fermionic evolution, in addition to the Majorana operators, for which $k = 1$.

Finally, it will be convenient to define the following Gaussian operation, which exchanges pairs of fermionic modes between qubits $j$ and $k$

\begin{align}
S_{jk} &= \frac{1}{2} \left(X_{j} Z^{\otimes (k - j) - 1} X_{k} + Y_{j} Z^{\otimes (k - j) - 1} Y_{k} + Z_j + Z_k \right) \nonumber \\
&= \frac{-i}{2} \left(c_{2j} c_{2k - 1} - c_{2j - 1} c_{2k} + c_{2j - 1} c_{2j} + c_{2k - 1} c_{2k}\right) \label{eq:fSWAP} \\
S_{jk} &= -i \exp\left[ \frac{\pi}{4}\left(c_{2j} c_{2k - 1} - c_{2j - 1} c_{2k} + c_{2j - 1} c_{2j} + c_{2k - 1} c_{2k}\right)\right] \nonumber
\end{align}

\noindent This operation effects $c_{2j - 1} \overset{S}{\leftrightarrow} c_{2k - 1}$ and $c_{2j} \overset{S}{\leftrightarrow} c_{2k}$ by conjugation.

It was shown, surprisingly, in Ref.~\cite{jozsa2008matchgates}, that Gaussian fermionic operations, together with 2-qubit nearest-neighbor $\mathrm{SWAP}$ operations, are universal for quantum computation. $\mathrm{SWAP}$ has a similar form to Eq.~(\ref{eq:fSWAP}), with the important distinction of a \emph{quartic} term in the Majorana operators

\begin{align}
&\mathrm{SWAP}_{j, j + 1} = \frac{1}{2} \left(I + X_j X_{j + 1} + Y_{j} Y_{j + 1} + Z_{j} Z_{j + 1}\right) \nonumber \\
&= \frac{1}{2} \left(I - i c_{2j} c_{2j + 1} + i c_{2j - 1} c_{2(j + 1)} - c_{2j - 1}c_{2j} c_{2j + 1} c_{2(j + 1)}\right) \mathrm{.} \label{eq:SWAPdef}
\end{align}

\noindent In contrast to Eq.~(\ref{eq:covariantommutator}), the quartic term maps between Majorana configuration operators of different degree under commutation, e.g.

\begin{align}
[Z_j Z_{j + 1}, c_{2j - 1}] = -2 c_{2j} c_{2j + 1} c_{2(j + 1)} \mathrm{,}
\label{eq:ZZcommutator}
\end{align} 

\noindent and in general 

\begin{align}
[Z_j Z_{j + 1}, C_{\va}] = 
\begin{cases} 2 (Z_j Z_{j + 1}) C_{\va} & \ |\va \cap \vec{q}_{j}| \ \mathrm{odd} \\
0 & \ |\va \cap \vec{q}_{j}| \ \mathrm{even}
\end{cases}\mathrm{,}
\end{align} 

\noindent where $\vec{q}_{j} \equiv (2j - 1, 2j, 2j + 1, 2j + 2)$. For $\va \subseteq \vec{q}_j$, $(Z_j Z_{j + 1}) C_{\va} = \pm C_{\vec{q}_{j}/\va}$, and $[Z_j Z_{j + 1}, c_k] = 0$ for $k \notin \vec{q}$. Since Eq.~(\ref{eq:ZZcommutator}) is analogous to a pair-production process for Majorana operators, we will refer to it as an ``interaction" between modes.

Finally, in Ref.~\cite{chapman2017classical}, it was shown that the infinite-temperature OTO correlator has an analytic closed-form expression when the unitary evolution is a Gaussian fermionic operation and $A \equiv i^a C_{\vec{\eta}}$ and $B \equiv i^b C_{\vec{\eta}}$ are Pauli operators (for integer $a$ and $b$), as 

\begin{align}
\mathcal{C}^2_{AB}(t) = \frac{1}{2} \left\{1 \pm \det{\left[\mathbf{u}_{\va [2n]} \left(\mathbf{I} - 2\mathbf{P}_{\ve} \right) \mathbf{u}^{\mathrm{T}}_{[2n] \va}\right]}\right\} \rm{,}
\label{eq:GFETheorem}
\end{align}

\noindent where the sign factor is simply $(-1)^{|\va| |\ve| + 1}$, and $\mathbf{P}_{\ve}$ is the projector onto the modes $\ve$ (i.e. it is diagonal with ones on the diagonal for modes in $\ve$ and zeroes elsewhere). In the following sections, we demonstrate a similar approach to that of Ref.s~\cite{xu2018accessing, sahu2018scrambling} to approximate the OTO correlator by considering interactions acting on only the fastest-traveling Majorana modes, near the lightcone edge, where we can effect the action of an interaction by an equivalent Gaussian fermionic transformation and apply Eq.~(\ref{eq:GFETheorem}).

\section{Results}
\label{sec:results}

\subsection{A. Universal Circuit Model}
\label{sec:ucmb}

\begin{figure}
\includegraphics[width=\columnwidth]{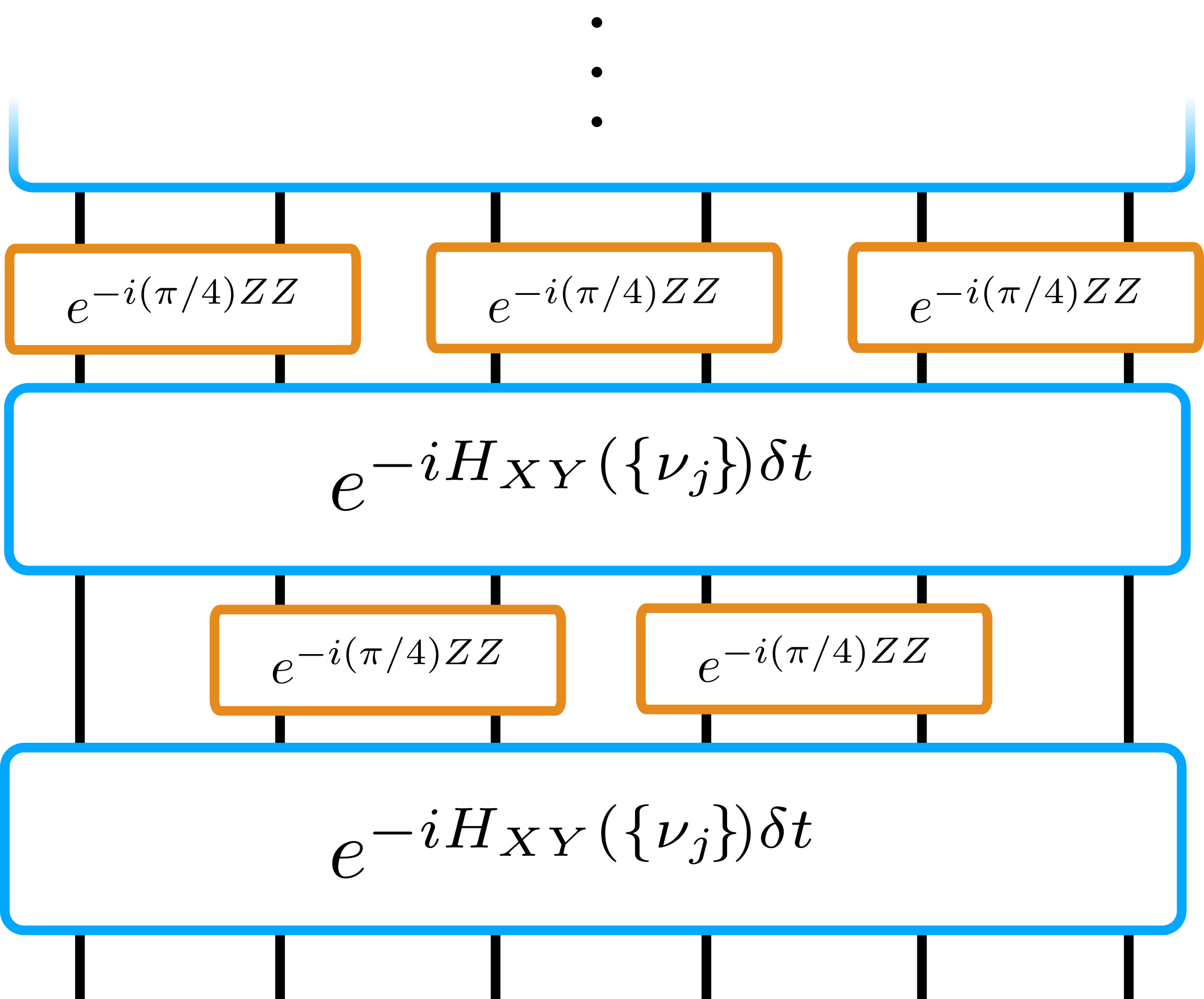}
\caption{(Color online) Our alternating circuit model of-interest, consisting of repeated layers of the form shown above. Global layers are periods of localizing Gaussian fermionic dynamics, with local disorder configuration $\{\nu_j\}$ and duration $\delta t$. Local gate layers consist of ``interaction gates,” which are non-Gaussian. The positions of these layers are alternated between qubits 1 and 2 being the left-most qubits to interact.}
\label{fig:alternatingcircuit}
\end{figure}

Our universal circuit model is shown in Fig.~\ref{fig:alternatingcircuit}. It consists of alternating layers between disordered Gaussian fermionic evolution and products of quartic fermion gates. The Gaussian fermionic evolution is given by

\begin{align}
&H_{XY}(\{\nu_j\}) = \sum_{j = 1}^{n - 1} \left(X_{j} X_{j + 1} + Y_{j} Y_{j + 1} \right) + \sum_{j = 1}^{n} \nu_j Z_j \nonumber \\
&\mbox{\hspace{5mm}}=  -i \sum_{j = 1}^{n - 1} \left(c_{2j} c_{2j + 1} - c_{2j - 1} c_{2(j + 1)}\right) - i \sum_{j = 1}^n \nu_j c_{2j - 1} c_{2j} 
\end{align}

\noindent with the $\{\nu_j\}$ a random local potential, chosen uniformly from the interval $[-\nu, \nu]$. We will demonstrate the existence of a many-body-localization transition to logarithmic scrambling at a disorder value of $\nu_c \sim 0.8$. When interactions are absent, however, propagation in this model is Anderson-localized for any nonzero disorder.

The quartic fermion ``interaction" gates are of the form

\begin{align}
\exp(-\frac{i\pi}{4} Z_j Z_{j + 1}) = \frac{1}{\sqrt{2}}(I - i Z_{j} Z_{j + 1})
\end{align}   

\noindent For $\va \subseteq \vec{q}_j$, we have 

\begin{align}
e^{\frac{i\pi}{4} Z_{j} Z_{j + 1}} C_{\va} e^{-\frac{i\pi}{4}Z_{j} Z_{j + 1}} &= (i Z_{j} Z_{j + 1})^{\left(|\va| \ \mathrm{mod} \ 2\right)} C_{\va} \mathrm{.}
\end{align}

\noindent It is crucial to our approximation that this gate be a Clifford operation (i.e. it preserves the set of Majorana configuration operators). This gate is equivalent to $\mathrm{SWAP}$ gate up to Gaussian fermionic gates, as 

\begin{figure*}
\includegraphics[width=1.9\columnwidth]{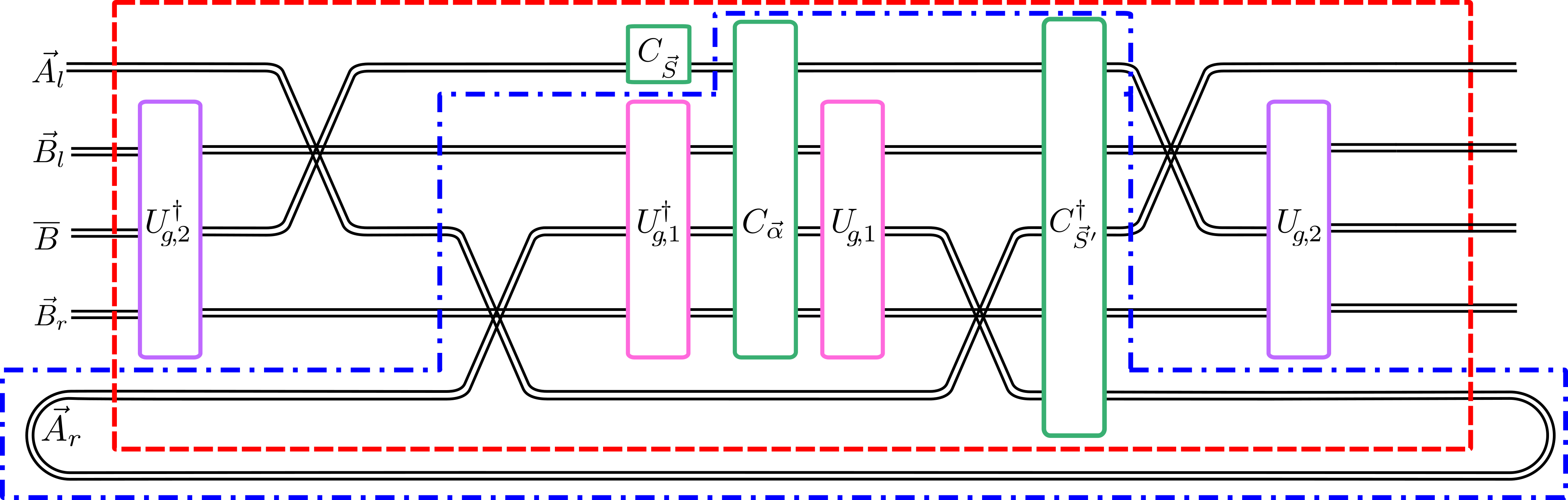}
\caption{(Color online) A Gaussian fermionic circuit whose evaluation is used to prove Eq.~(\ref{eq:modCBform}). The circuit is depicted in the qubit picture, with pairs of circuit wires representing collections of arbitrarily many qubits. As such, all Majorana configuration operators ($C_{\vec{\alpha}}$, $C_{\vec{S}}$, and $C_{\vec{S}^{\prime}}^{\dagger}$) are drawn with support all the way up to the top qubit, since there may in principle be a string of Pauli-$Z$ operators up to this qubit if the number of Majorana operators in the configuration is odd. Nevertheless, we do assume that $\vec{S}^{\prime} \subseteq \vec{A}_r$. That is, if $|\vec{S}^{\prime}|$ is odd, then $C^{\dagger}_{\vec{S}^{\prime}}$ only acts as a string of $Z$ operators on the qubits corresponding to modes $(\vec{A}_l, \vec{B}_l, \overline{B}, \vec{B}_r)$, and so all Gaussian fermionic gates commute with $C^{\dagger}_{\vec{S}^{\prime}}$ on these qubits. Crossing lines represent the Gaussian fermionic operation of rearranging subsets of fermionic modes (i.e. products of $\{S_{jk}\}$ defined in the text) while preserving these subsets' internal ordering, but possibly applying phases. Similarly, $U_{g, 1}$ and $U_{g, 2}$ are Gaussian fermionic unitaries. The self-contracted wires at the bottom represent a partial trace over this subset of qubits (notice we can reliably trace over the last subset since no Majorana configurations will have support here). The dotted and dot-dashed boxes indicate which portion of the circuit to contract in one step during its evaluation, followed by the contraction over everything else in the second step. For example, the dot-dashed-box contraction consists of conjugating $C_{\vec{\alpha}}$ by $U_{g, 1}$, exchanging modes $\overline{B}$ and $\vec{A}_r$, and tracing over $\vec{A}_r$ with respect to $C_{\vec{S}^{\prime}}^{\dagger}$ in the first step. Next, the modes  $\overline{B}$ and $\vec{A}_l$ are exchanged and $U_{g, 2}$ is applied in the second step. This contraction partitioning corresponds to the left-hand-side of Eq.~(\ref{eq:modCBform}). The dotted box consists of applying all Gaussian fermionic unitaries in the first step, and then performing the partial trace with respect to $C_{\vec{S}^{\prime}}^{\dagger}$ in the second step (remember, $C_{\vec{S}^{\prime}}^{\dagger}$ commutes with Gaussian fermionic unitaries on modes outside of $\vec{A}_r$). This contraction partitioning corresponds to the right-hand-side of Eq.~(\ref{eq:modCBform}). As these two contractions must evaluate to the same operator, their equivalence implies the equality.}
\label{fig:modCBcircuit}
\end{figure*}

\begin{align}
&e^{\frac{i\pi}{4}} \mathrm{SWAP} = \exp{\left(-\frac{i\pi}{4} Z_{j} Z_{j + 1} \right)} \nonumber \\ 
&\mbox{\hspace{20mm}} \times \exp{\left[\frac{5i\pi}{4}\left(Z_j + Z_{j + 1}\right)\right]} S_{j, j + 1} \mathrm{.}
\end{align}

\noindent Thus, the inclusion of these gates extends Gaussian fermionic operations to computationally universal quantum circuits.

\subsection{Exact Formula}

In Appendices~\hyperref[sec:mcbformuladproof]{A} and \hyperref[sec:mcbformula]{B}, we prove an exact formula for a general quantum circuit expressed as a product of Gaussian fermionic evolution and interaction gates. This formula follows from a modification to the Cauchy-Binet formula (\ref{eq:CBformula}):

\begin{align}
&\sum_{\vb \subseteq \vec{B} \equiv \left(\vec{B}_l, \vec{B}_r \right)} \det\left[\left(\mathbf{u}_1\right)_{\va, \vb \cup \vec{S}^{\p}}\right] \det\left[\left(\mathbf{u}_2\right)_{\vb \cup \vec{S}, \vg}\right] \nonumber \\ 
& \mbox{\hspace{5mm}} = (-1)^{|\vec{S}| |\vec{S}^{\p}|} \det
\begin{bmatrix}
\mathbf{0}_{|\vS| \times |\vS^{\p}|} & \left(\mathbf{u}_2\right)_{\vS \vg} \\
\left(\mathbf{u}_1\right)_{\va \vS^{\p}} & \left(\mathbf{u}_1\right)_{\va \vec{B}} \left(\widetilde{\mathbf{u}}_2\right)_{\vec{B} \vg}
\end{bmatrix} \rm{.}
\label{eq:modCBform}
\end{align}

\noindent Letting $\overline{B}$ be the set-complement of $\left(\vec{A}_{l}, \vec{B}_l, \vec{B}_r, \vec{A}_r \right)$ in the set of all modes (see Fig.~\ref{fig:modCBcircuit}), $\vec{S}$, $\vec{S}^{\p} \subseteq \overline{B}$ are fixed sets of modes which are not summed over. $\vec{B}_{l}$ is a contiguous subset of modes to the ``left" of and disjoint from $\overline{B}$, and $\vec{B}_{r}$ is a contiguous subset of modes to the ``right" of and disjoint from $\overline{B}$. Furthermore,

\begin{align}
\widetilde{\mathbf{u}}_2 \equiv \left(\mathbf{I} - 2\delta_{|\vec{S}| + |\vec{S}^{\prime}| = 1 \left(\mathrm{mod} \ 2\right)}\mathbf{P}_{\vec{B}_l}\right)\mathbf{u}_2 \mathrm{.}
\label{eq:utdef} 
\end{align}

\noindent That is, $\widetilde{\mathbf{u}}_2 = \mathbf{u}_2$ unless $|\vec{S}|$ and $|\vec{S}^{\prime}|$ have opposite parity, in which case the rows of $\mathbf{u}_{2}$ corresponding to the modes $\vec{B}_l$ are multiplied by $(-1)$ to obtain $\widetilde{\mathbf{u}}_2$. Eq.~(\ref{eq:modCBform}) was proved for the special case where $\vec{S} = \vec{S}^{\prime}$ in Ref.~\cite{chapman2017classical}. 

Though we prove Eq.~(\ref{eq:modCBform}) rigorously in Appendix~\hyperref[sec:mcbformula]{B} using properties of determinants, a simple pictorial proof can be seen in Fig. \ref{fig:modCBcircuit} (for details of this version of the proof, see Appendix \hyperref[sec:mcbformuladproof]{A}). This figure depicts a circuit consisting of fixed evolution by Gaussian fermionic unitaries $U_{g, 1}$, $U_{g, 2}$, and rearrangements of fermionic modes (the crossing wires, which are products of the $\{S_{jk}\}$) with the ancillary modes ($\vec{A}_{l}$, $\vec{A}_r$). The self-contracted wires at the bottom represent a partial trace taken on the last subset of qubits. The modified Cauchy-Binet formula, Eq.~(\ref{eq:modCBform}), follows from considering the two equivalent ways we can choose to evaluate this circuit: either we contract everything in the dotted box and perform the partial trace afterward, or we contract everything in the dot-dashed box (including taking the partial trace) and perform the remainder of the Gaussian fermionic evolution afterward. These two different contraction orderings yield the same operator (since they are the same circuit), yet the former evaluates to the right-hand-side of Eq.~(\ref{eq:modCBform}), and the latter evaluates to the left-hand-side. See the caption under Fig.~\ref{fig:modCBcircuit} or Appendix~\hyperref[sec:mcbformula]{A} for details.

\begin{figure}
\includegraphics[width=\columnwidth]{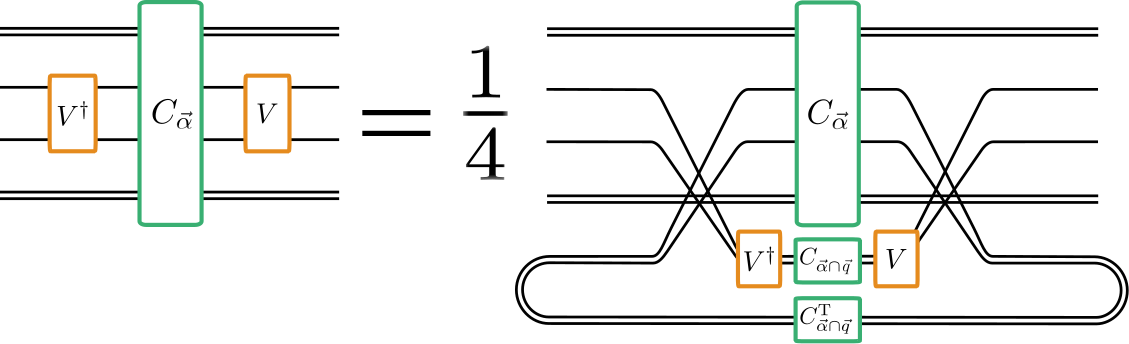}
\caption{(Color online) The action of an interaction gate $V \equiv e^{-i \left(\pi/4\right) Z Z}$, acting on the modes $\vec{q}$ can be reproduced by exchanging the modes $\vec{q}$ with an ancilla occupied by $V^{\dagger} C_{\va \cap \vec{q}} V$, and taking the partial trace with respect to $C_{\va \cap \vec{q}}$. Note that, since $C^{\mathrm{T}}_{\va \cap \vec{q}}$ potentially only differs from $C_{\va \cap \vec{q}}$ by a sign, the product $V^{\dagger} C_{\va \cap \vec{q}} V C^{\mathrm{T}}_{\va \cap \vec{q}}$ does not have support on the qubits above it due to the parity-preserving property of $V$. Choosing the dotted contraction-ordering in Fig.~\ref{fig:modCBcircuit} above, whereby the partial trace is taken at the very end, we can iterate the application of this identity and compute the OTO correlator as the single trace of a Majorana configuration under Gaussian fermionic evolution, with respect to the appropriate Majorana configuration output by the interaction gates.}
\label{fig:interactionidentityexact}
\end{figure}

We construct an exact formula for the OTO correlator of universal quantum circuit dynamics by iteratively applying Eq.~(\ref{eq:modCBform}). Since our interaction gate is parity-preserving, Eq.~(\ref{eq:modCBform}) can be realized as the identity shown in Fig.~\ref{fig:interactionidentityexact}, whereby the input modes to the interaction are exchanged with the appropriate output modes on an ancilla, and the ancilla is traced over. Choosing the contraction corresponding to the dotted box in Fig.~\ref{fig:modCBcircuit} (i.e. performing all traces at the end) we can calculate the OTO correlator as a series of terms of a similar form to Eq.~(\ref{eq:GFETheorem}), summed over all exponentially-many inputs to each interaction gate. Each such input configuration can be thought of as a particular ``computational path" in the operator space of Majorana configurations, and the OTO correlator is realized as a superposition over all of these paths, which will interfere in general. Our algorithm for exactly calculating the OTO correlator then proceeds as follows: \\

\noindent \textbf{Exact series for the OTO correlator:} \\ \\
\noindent \emph{Given:} \begin{enumerate}
\item Universal quantum circuit $U \equiv U_1 U_2 \dots U_N$ on $n$ qubits with $g$ interaction gates. \\
\item Pauli observables $A \equiv i^a C_{\vec{\eta}}$ and $B \equiv i^b C_{\vec{\alpha}}$, for integers $a$ and $b$.\\
\end{enumerate}

\begin{figure}
\includegraphics[width=\columnwidth]{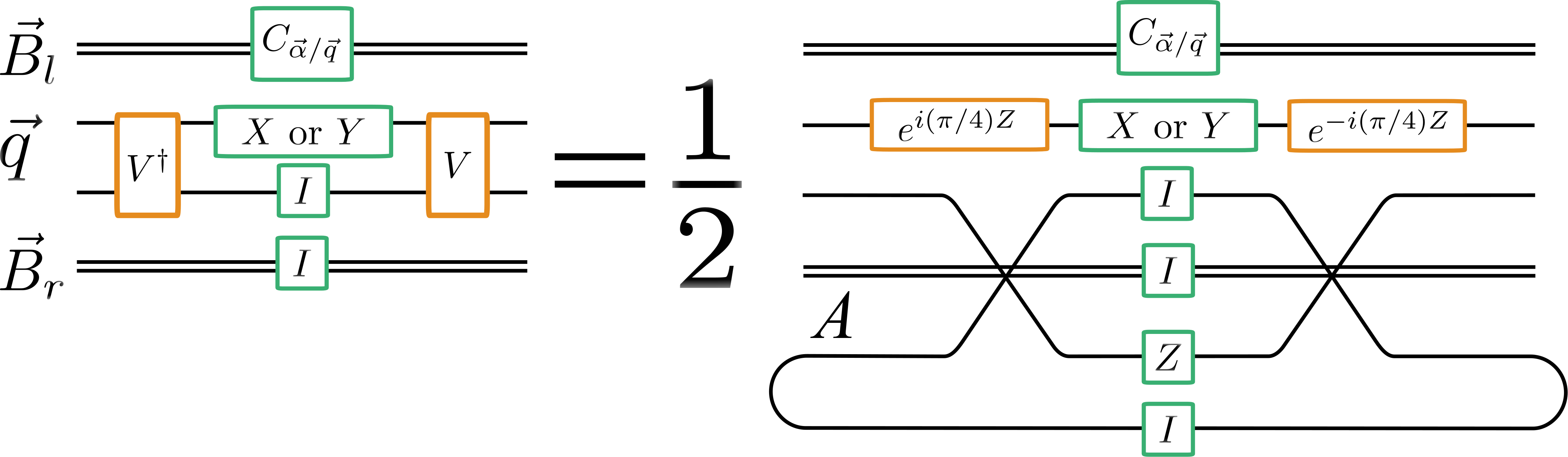}
\caption{(Color online) We approximate the action of an interaction gate on the lightcone using the identity shown above (a similar identity holds when the identity input is replaced with a $Z$). Namely, the effect of conjugating an interaction gate on a single Pauli $X$ or $Y$ is equivalent (up to normalization) to performing a local $Z$ rotation on that qubit, together with exchanging (via a Gaussian operation $S_{jk}$) a Pauli $Z$ with an ancilla $A$, which is then traced over.}
\label{fig:interactionidentityapprox}
\end{figure}

\noindent \emph{Construct:} An orthogonal, symmetric matrix $\mathbf{K}$ from which the infinite-temperature OTO correlator $\mathcal{C}_{AB}(t)^2$ can be calculated as a sum of minors from $\mathbf{K}$, as

\begin{align}
&1 - 2 \mathcal{C}_{AB}(t)^2 \nonumber \\ & \nonumber \\
&= \sum_{\substack{\vec{B} = \bigcup_{i = 1}^g \vb_i \subseteq \vec{A}_{(N)} \\
\vec{B}^{\p} = \bigcup_{i = 1}^g \vb^{\p}_i \subseteq \vec{A}^{\p}_{(N)}}} (-1)^{f(\vec{B}, \vec{B}^{\p})} \det\left[ \mathbf{K}_{\left[\vec{B}^{\p}, \mathcal{V} \left(\vec{B}\right), \va \right] \left[\vec{B}, \mathcal{V} \left(\vec{B}^{\p} \right), \va \right]}\right]
\label{eq:thm1eq1}
\end{align}

\noindent for ancillary modes, $\vec{A}_{(N)}$, $\vec{A}^{\prime}_{(N)}$, integers $\{f(\vec{B}, \vec{B}^{\prime})\}$, and $\mathcal{V}$ the map relating the Majorana configuration-tuple input to a set of interaction gates $V^{\otimes g}$ to the configuration-tuple of the output. 

\begin{enumerate}
\item Let $\mathbf{u}_{(0)} \equiv \mathbf{I}_{2n}$, the $2n \times 2n$ identity matrix, and $\vec{A}_{(0)} \equiv ()$, the empty tuple. \\
\item For $j \in (1, \dots, N)$: \\
\begin{enumerate} 
\item If $U_j$ is Gaussian fermionic, corresponding to $\mathbf{u} \in \mathrm{SO}(2n)$ \\
\begin{enumerate}
\item $\mathbf{u}_{(j)} \equiv \mathbf{u}_{(j - 1)} \left(\mathbf{I}_{\vec{A}_{(j - 1)}} \oplus \mathbf{u} \right)$ \\
\item $\vec{A}_{(j)} \equiv \vec{A}_{(j - 1)}$ 
\end{enumerate}
\item If $U_j = \exp{\left(-\frac{i \pi}{4} Z_{i_j} Z_{i_j + 1}\right)}$, an interaction gate between qubits $(i_j, i_j + 1)$, on modes $\vec{q}_{j} \equiv \left(2 i_j - 1, 2 i_j, 2 i_j + 1, 2 i_j + 2 \right)$
\begin{enumerate}
\item Let $\vec{B} \equiv \left(\vec{A}_{(j - 1)}, [2n] + |\vec{A}_{(j - 1)}|\right)$, where addition indicates adding a fixed value to every index of the set. Let $\widetilde{q}_j \equiv \vec{q}_j + |\vec{A}_{(j - 1)}|$ and $\overline{q}_j \equiv \vec{B}/\widetilde{q}_j$
\item $\mathbf{u}_{(j)} \equiv \begin{bmatrix} \mathbf{0}_{4 \times 4} & \mathbf{I}_{\widetilde{q}_j \vec{B}} \\ \left(\mathbf{u}_{(j - 1)}\right)_{\vec{B}, \widetilde{q}_j} & \left(\mathbf{u}_{(j - 1)}\right)_{\vec{B}, \overline{B}}\mathbf{I}_{\overline{B}, \vec{B}} \end{bmatrix}$ 
\item $\vec{A}_{(j)} \equiv ([4], \vec{A}_{(j - 1)} + 4)$
\end{enumerate}
\end{enumerate}
\item \begin{enumerate}
\item Let $\vec{p}\equiv  [2n] + |\vec{A}_{(j)}|$, $\vec{B} \equiv \left(\vec{A}_{(j)}, \vec{p}\right)$
\item Let $\mathbf{R} \equiv (-1)^{|\vec{\eta}|} \left(\mathbf{I}_{2n} - 2\mathbf{P}_{\vec{\eta}}\right)$, a $2n \times 2n$ diagonal matrix whose diagonal elements are $(-1)^{|\vec{\eta}| + 1}$ for modes in $\vec{\eta}$ and $(-1)^{|\vec{\eta}|}$ otherwise.
\item $\mathbf{K} \equiv \begin{bmatrix}
\mathbf{0}_{|\vec{A}_{(j)}| \times |\vec{A}_{(j)}|} & \left(\mathbf{u}^{\mathrm{T}}_{(j)}\right)_{\vec{A}_{(j)}, \vec{B}} \\  
\left(\mathbf{u}_{(j)}\right)_{\vec{B}, \vec{A}_{(j)}} & \left(\mathbf{u}_{(j)}\right)_{\vec{B}, \vec{p}}\mathbf{R} \left(\mathbf{u}^{\mathrm{T}}_{(j)}\right)_{\vec{p}, \vec{B}}
\end{bmatrix}$
\end{enumerate}
\end{enumerate}

\noindent The phases $(-1)^{f(\vec{B}, \vec{B}^{\p})}$ are calculated by iterated application of Eq.~(\ref{eq:modCBform}) every time an interaction is applied in step 2(b) and are calculated explicitly in Appendix~\hyperref[sec:exactoto]{C}.

\subsection{Approximative method}

Though our formula in Eq.~(\ref{eq:thm1eq1}) is exact, the number of terms in the sum scales exponentially in the number $g$ of interaction gates (though each term can be evaluated using only polynomial resources in the number of qubits). We therefore make the physical restriction to the fastest traveling modes, which allows us to approximate the lightcone envelope by a series truncated to a single determinant corresponding to an effective Gaussian fermionic evolution. Our workhorse identity is shown graphically in Fig. \ref{fig:interactionidentityapprox} and given by: \\

\begin{figure}
\includegraphics[width=0.95\columnwidth]{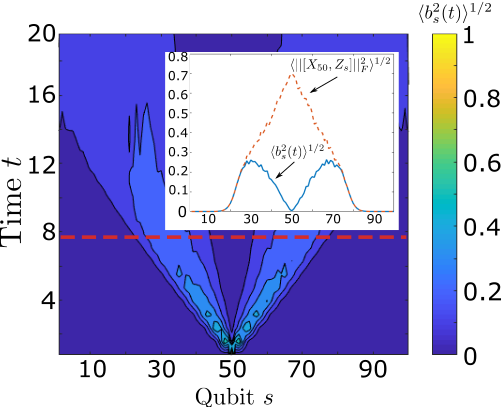}
\caption{(Color online) The lightcone boundary as calculated by Eq.~(\ref{eq:Bdef}) for Gaussian fermionic evolution (no interaction gates) on $n = 100$ qubits and disorder strength $\nu = 1$, averaged over 25 disorder realizations, with $\delta t = \pi/4$. We see that the boundary spreads, leaving a depletion region in the center, and becomes wider with time. (Inset) The boundary at the time slice indicated by the dotted line, at $t = \frac{5\pi}{2}$, together with $\mathcal{C}_{XZ}$ at the same time. We see clearly a region where the lightcone has very nearly the exact value of the boundary, indicating that our approximate method becomes exact in this region, since the underlying assumption to the approximation is perfectly satisfied here.}
\label{fig:boundarycone}
\end{figure}

\noindent \textbf{Conditional Gaussian evolution:} For a Majorana operator $c_{\mu}$, with $\mu \in (2j - 1, 2j)$, we have
\begin{align}
&e^{\frac{i\pi}{4} Z_{j} Z_{j + 1}} c_\mu e^{\frac{-i\pi}{4} Z_{j} Z_{j + 1}} \nonumber \\
&\mbox{\hspace{5mm}}= \frac{1}{2}\tr_A\left[\left(e^{\frac{i\pi}{4} Z_{j}} c_\mu e^{\frac{-i\pi}{4} Z_{j}}\right) \left(S^{\dg}_{j + 1, A} Z_A S_{j + 1, A}\right)\right] \rm{,} \label{eq:rlightcone} \\
&e^{\frac{i\pi}{4} Z_{j} Z_{j + 1}} \left(c_\mu Z_{j + 1}\right) e^{\frac{-i\pi}{4} Z_{j} Z_{j + 1}} \nonumber \\
&\mbox{\hspace{5mm}}= \frac{1}{2}\tr_A\left[\left(e^{\frac{i\pi}{4} Z_{j}} c_\mu e^{\frac{-i\pi}{4} Z_{j}}\right) \left(S^{\dg}_{j + 1, A} Z_{j + 1} S_{j + 1, A}\right) Z_{A}\right] \label{eq:llightcone}
\end{align}
Similar identities hold for $\mu \in (2j + 1, 2j + 2)$.
\label{lem:cme}

The identities above relate the action of an interaction gate on an operator at the lightcone edge to that of a corresponding equivalent Gaussian fermionic gate, allowing us to approximately simulate it classically. By similar logic to that argued in Ref.s~\cite{xu2018accessing, sahu2018scrambling}, we expect the propagation of any error introduced in this approximation to be bounded by the speed of light of the underlying dynamics. Such error results from terms in the operator expansion of $B(t)$ which are not of the form shown in Eq.s~(\ref{eq:rlightcone}) or (\ref{eq:llightcone}) (or the corresponding form for $\mu \in (2j + 1, 2j + 2)$). We calculate the weight of the assumption-satisfying terms using the modified Cauchy-Binet formula (\ref{eq:CBformula}). Let 
\begin{align}
(-i)^b B(t) = \sum_{\vb} \det \left(\mathbf{u}_{\va \vb} \right) C_{\vb}
\label{eq:Bdef}
\end{align}

\noindent That is, assume $B = i^b C_{\va}$ and that the evolution up to time $t$ is described by a Gaussian fermionic operation as in Eq.~(\ref{eq:gfevolution}). Additionally, let 

\begin{align}
b_s^2(t) \equiv 
\begin{cases}
\sum_{\substack{\vb \ \mathrm{with} \ X_s I^{\otimes (n - s)} \ \\ \mathrm{or} \ Y_s I^{\otimes (n - s)} \ \mathrm{present}}} \det \left(\mathbf{u}_{\va \vb} \right)^2 & (s \geq \lfloor n/2 \rfloor) \\
\sum_{\substack{\vb \ \mathrm{with} \ I^{\otimes (s - 1)} X_s \ \\ \mathrm{or} \ I^{\otimes (s - 1)} Y_s \ \mathrm{present}}} \det \left(\mathbf{u}_{\va \vb} \right)^2 & (s \leq \lfloor n/2 \rfloor)\end{cases} \mathrm{.}
\label{eq:boundarydef}
\end{align}

\noindent This is the total weight of the terms which do not commute with $\exp{\left(\frac{i\pi}{4}Z_{s} Z_{s \pm 1}\right)}$ and which correspond to the right or left lightcone edge being found at qubit $s$. The condition in the upper sum (for which $s \geq \lfloor n/2 \rfloor$) will only be met if there exists a tuple $\vb^{\prime}$ for which $\vb = (\vb^{\prime}, 2s - 1)$ or $\vb = (\vb^{\prime}, 2s)$. Similarly, for $B \equiv X_{\lfloor n/2 \rfloor}$, the condition in the lower sum (for which $s \leq \lfloor n/2 \rfloor$) will only be met if there exists a tuple $\vb^{\prime}$ for which $\vb = ([2s - 2], 2s - 1, \vb^{\prime})$ or $\vb = ([2s - 2], 2s, \vb^{\prime})$, where $[2s - 2] = (1, 2, \dots, 2s - 2)$ (since we know the total number of modes in $\vec{\beta}$ must be odd for $B = X_{\lfloor n/2 \rfloor}$). We therefore apply the modified Cauchy-Binet formula, Eq.~(\ref{eq:modCBform}), to calculate this quantity exactly (see Appendix~\hyperref[sec:exactboundary]{D} for the full expression). For an illustration of the efficacy of this measure for the boundary, see Fig.~\ref{fig:boundarycone}.

We utilize this quantity $b_s(t)$, together with our conditional Gaussian evolution identities Eq.s~(\ref{eq:rlightcone}), (\ref{eq:llightcone}) in our approximation algorithm for the OTO correlator, which proceeds as follows: \\ 

\noindent \textbf{Approximation to Interaction by Conditional Gaussian Evolution:} \\ \\
\noindent \emph{Given:} $B(t - \delta t)$, described by an orthogonal matrix $\mathbf{u}$ as in Eq.~(\ref{eq:Bdef}) and a global tolerance $\varepsilon$ \\ \\

\noindent \emph{Approximate:} $V_j B(t - \delta t) V_j^{\dg}$ for $V_j = \exp\left(-\frac{i\pi}{4}Z_j Z_{j + 1}\right)$ \\

\begin{enumerate}
\item Calculate $b_s(t)$ for all $s \in (1, \dots, 2n)$. \\
\item For $s \in (1, \dots, 2n)$, \emph{if} $b_s(t) \geq \varepsilon$: \\ \\
\emph{If} $s \geq \lfloor n/2 \rfloor$ and $j = s$, \\
\emph{or} $s \leq \lfloor n/2 \rfloor$ and $j = s - 1$: \\ \\
Replace $V_j$ with a Gaussian operation by Eq.~(\ref{eq:rlightcone}) for $\mu \in (2s - 1, 2s)$.
\end{enumerate}

\noindent Each approximation step introduces an extra ancillary qubit (see Fig. \ref{fig:interactionidentityapprox}), but once again, we can perform the trace over the entire ancillary system as one with the trace in Eq.~(\ref{eq:cfrelation}) (notice that the normalization is kept consistent as we add each ancillary qubit). This allows us to straightforwardly apply Eq.~(\ref{eq:GFETheorem}) to calculate the OTO correlator.

\begin{figure}[t]
\includegraphics[width=0.95\columnwidth]{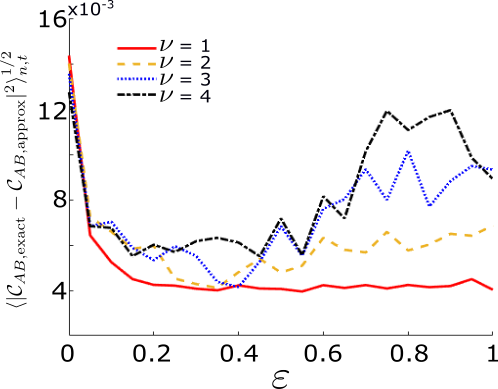}
\caption{(Color online) Average (per pixel) Frobenius-norm error between the exact and approximate lightcones on 6 qubits for disorder values $\nu\in\{1, 2, 3, 4\}$ (see Fig.~\ref{fig:comparison} for an example at disorder $\nu = 10$) as a function of the variationally optimized free parameter $\varepsilon$, averaged over 25 samples. We see the clear emergence of a local minimum at $\varepsilon\in [0.2, 0.4]$ as the disorder is increased above $\nu\sim 1$. This is due to the fact that, when the Gaussian fermionic evolution is nearly delocalized, the decision of whether to keep an interaction gate makes negligible difference for small system size, since interaction gates cannot extend the lightcone beyond a ballistic profile.}
\label{fig:frobdiffsnu}
\end{figure}

\begin{figure*}
\includegraphics[width=2\columnwidth]{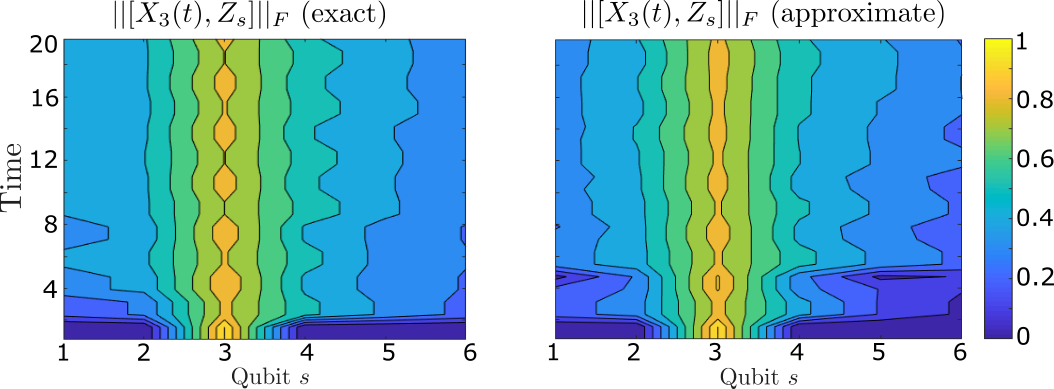}
\caption{(Color online) A comparison between the output of our algorithm and a brute-force calculation done on $n = 6$ qubits with disorder $\nu = 10$, $\varepsilon = 0.2$, $\delta t = \pi/4$, averaged over $50$ disorder realizations. We see good agreement at the interior of the lightcone, though edge fluctuations become more prominent under the approximation.}
\label{fig:comparison}
\end{figure*}

\begin{figure*}
\includegraphics[width=2\columnwidth]{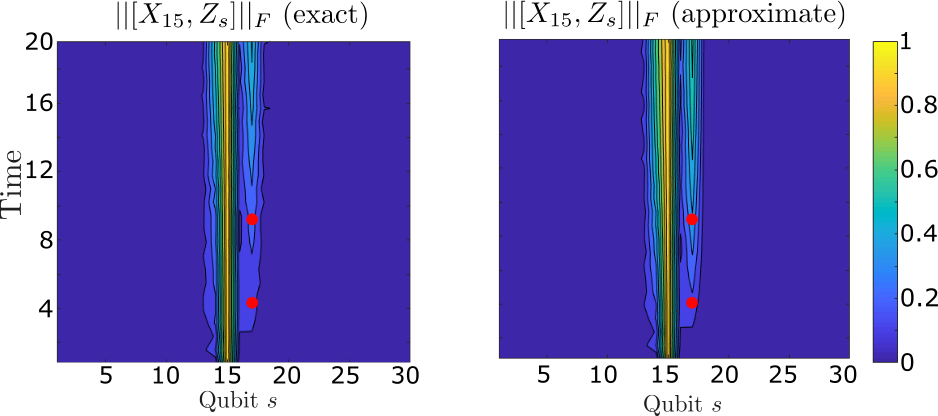}
\caption{(Color online) A comparison between the output of our algorithm and an exact calculation, which scales exponentially in the number of interaction gates, but efficiently in the number of qubits, for disorder value $\nu = 10$, $\delta t = \pi/4$, and two interaction gates at qubits $(17, 18)$ at times $t = \frac{3\pi}{2}$, $3\pi$ (red circles). We consider only a single disorder instance here, so edge fluctuations are more pronounced. We nevertheless observe good agreement between the envelopes of the two lightcones at the optimized value $\varepsilon = 0.2$.}
\label{fig:2swap}
\end{figure*}

\subsection{Variational Optimization of the Free Paramater}

The free parameter $\varepsilon$ in our algorithm effectively decides where we would like to truncate the free-particle lightcone. Since the lightcone edge will actually have some finite width (related to the decay length $\eta$ in the bounds Eq.s~(\ref{eq:LRbound}) and (\ref{eq:loglightcone})), this free parameter is necessary. A key assumption of our algorithm is that errors introduced \emph{inside} the lightcone envelope will not change the propagation of the envelope itself, since such errors cannot travel faster than the speed of light. It is therefore important that we capture this lightcone edge precisely, without applying our approximation to interactions that fall outside of the lightcone of the exact dynamics. We are able to remove the free parameter by variationally optimizing the Frobenius norm of our approximate lightcone relative to the exact, brute-force calculation for small system size. In Fig.~\ref{fig:frobdiffsnu}, we demonstrate the emergence of a local minimum in the average-case Frobenius norm error between our approximation at given $\varepsilon$ and the brute-force calculation for $n = 6$ qubits, as a function of $\varepsilon$, as we tune the disorder strength from $\nu = 1$ to $\nu = 4$. We attribute this the appearance of this local minimum to the fact that, at low disorder, we expect the decision of whether to keep a given interaction gate to be less important, since an interaction cannot extend the lightcone beyond a ballistic one. The appearance of a local minimum is therefore consistent with the emergence of a genuine many-body-localization transition. In Fig.~\ref{fig:comparison}, we compare the output of our algorithm to that of a brute-force calculation at $\varepsilon = 0.2$, the optimal value, for $n = 6$, averaged over $50$ disorder realizations of strength $\nu = 10$, where we observe good agreement (we choose high disorder here so that features of the lightcone can be seen within a region 6 qubits wide). In Fig.~\ref{fig:2swap}, we examine the correctness of our algorithm in the opposite extreme, where the number of qubits is large ($n = 30$) and the number of interaction gates is limited to two (at the red circles), using our exact formula Eq.~(\ref{eq:thm1eq1}). We see excellent agreement between the lightcone envelopes at disorder strength $\nu = 10$, again at the optimized value of $\varepsilon = 0.2$.

\begin{figure*}
\includegraphics[width=1.9\columnwidth]{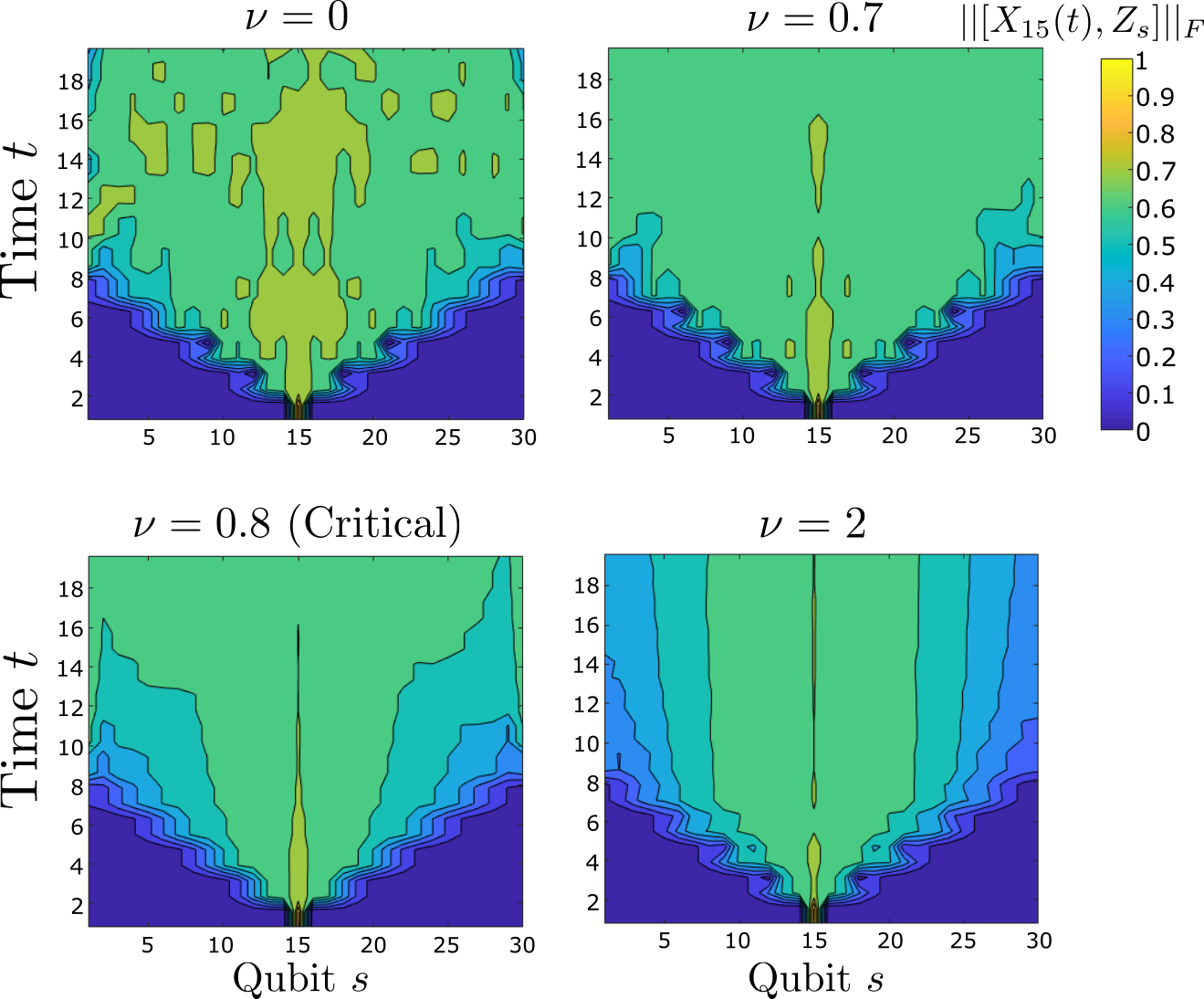}
\caption{(Color online) Lightcone transition from ballistic propagation at low disorder ($\nu < \nu_c \approx 0.8$) to logarithmically localized propagation at high disorder ($\nu > \nu_c$), averaged over $10^3$ disorder realizations. A characteristic feature of the localized phase is a region where the OTO correlator is maximized ($> 0.6$) about the location of the initial excitation.}
\label{fig:alternatinglightcone}
\end{figure*}

\begin{figure}
\includegraphics[width=0.95\columnwidth]{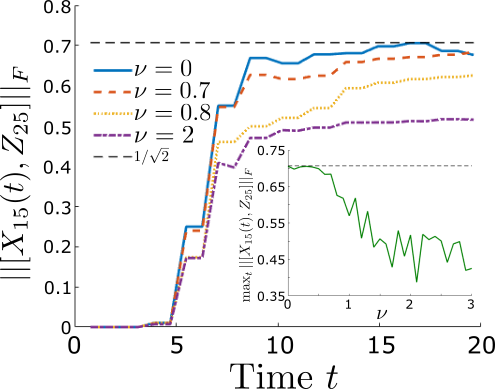}
\caption{(Color online) The OTO correlator value at a fixed qubit $(s = 25)$ for the four disorder values shown in Fig.~\ref{fig:alternatinglightcone}. We see that below the critical value, the OTO correlator approaches a limiting value of $1/\sqrt{2}$, the Page scrambled value (see main text), and a lower limiting value in the localized regime. We plot this limiting value as a function of disorder strength (inset), where we see a clear deviation from Page scrambling for disorder values $\nu \geq \nu_c \approx 0.8$.}
\label{fig:asympval}
\end{figure}

\begin{figure}
\includegraphics[width=0.95\columnwidth]{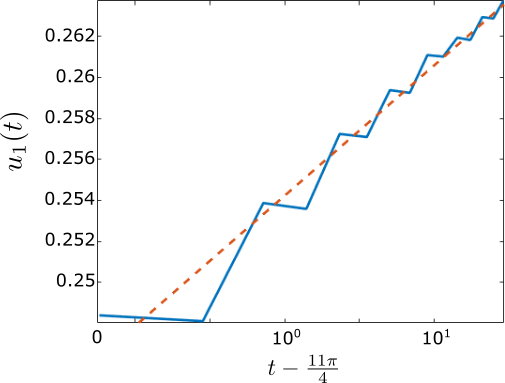}
\caption{(Color online) The principal singular vector, $u_{1}(t)$, of the lightcone in Fig.~\ref{fig:alternatinglightcone} for $\nu = 2$, treated as a numerical matrix, and plotted on a logarithmic (base $10$) $x$-axis. Logarithmic scrambling is observed by this method for $t \geq \frac{11\pi}{4}$, as prior to this, the principal-singular-vector behavior is dominated by a ballistically-spreading low-amplitude component.}
\label{fig:loglightsvd}
\end{figure}

\section{Many-body location transition}

Our main numerical result is shown in Fig.s \ref{fig:alternatinglightcone} and \ref{fig:asympval}, where we demonstrate that our universal circuit model, consisting of alternating disordered Gaussian fermionic evolution and interaction gates as in Fig.~\ref{fig:alternatingcircuit}, exhibits a many-body-localization transition in $\mathcal{C}_{XZ}(t)$ as we tune the disorder strength across a critical value $\nu_c \approx 0.8$. In Fig. \ref{fig:alternatinglightcone}, we plot the lightcone propagation for disorder values $\nu \in \{0, 0.7, 0.8, 2\}$ across the critical disorder strength for $10^3$ samples of the disorder. We note a clear emergence of a highly localized region of maximal value ($\mathcal{C}_{XZ} \approx 0.7$), which persists for all time in this figure when the disorder strength is greater than $\nu_c$. This is approximately the operator Page-scrambled value of $\frac{1}{\sqrt{2}}$ \cite{sekino2008fast}, where the operator $X_{15}(t)$ has equal weight for all four possible Pauli operators $\{I_s, X_s, Y_s, Z_s\}$ at a given site $s$. That is, contracting $X_{15}(t)$ on one side of the thermofield double state Eq.~(\ref{eq:thermofielddoubledef}) and tracing over all but qubit $s$ and $s^{\prime}$ on subsystems $r$ and $\tilde{r}$, respectively, would give the 2-qubit maximally mixed state $\frac{1}{4} I$. Commutation with $Z_s$ keeps only the weights on $\{X_s, Y_s\}$, each of which are $\frac{1}{4}$. Adding these and taking the square root gives the value of $\mathcal{C}_{XZ}$ to be approximately $\frac{1}{\sqrt{2}}$. Our numerics are therefore consistent with the fact that, within the localized region, the operator $X_s$ is approximately Page scrambled. Since this property is preserved under Clifford-gate evolution, such as by our interaction gate, the existence of this Page-scrambled region justifies our approximation to neglect the action of such gates acting \emph{inside} the lightcone, since they would have negligible effect on the lightcone interior.

In Fig.~\ref{fig:asympval}, we plot a spatial slice of each of the lightcones in Fig. \ref{fig:alternatinglightcone} at $s = 25$. We see that below the critical value of $\nu_c = 0.8$, the limiting value is very nearly the Page value $1/\sqrt{2}$, while above the critical value, it begins to decrease with $\nu$. In the inset, we plot the limiting value (which we take as the maximum) as a function of disorder strength, where we see that it clearly begins to deviate strongly from the Page value as we increase the disorder past the critical value. In Fig.~\ref{fig:loglightsvd}, we plot the principal temporal singular vector of the $\nu = 2$ lightcone in Fig. \ref{fig:alternatinglightcone}, treated as a numerical matrix, against a logarithmic $x$-axis for $t \geq \frac{11\pi}{4}$. The principal singular component of this matrix is the closest product approximation to the lightcone in Frobenius norm, and so this provides a robust, numerically inexpensive means of analyzing the dynamical phase (see Appendix G in Ref.~\cite{chapman2017classical} for details). Prior to $t = \frac{11\pi}{4}$, this principal vector is dominated by a ballistically-spreading low-amplitude component (see Fig. \ref{fig:alternatinglightcone}), but for $t \geq \frac{11\pi}{4}$, we see the OTO correlator growth is linear on this semi-logarithmic plot, indicating that the lightcone is logarithmic after this time. We choose to neglect this early-time behavior since we are primarily interested in the long-time asymptotic growth of the OTO correlator for our model.

\section{Discussion}

We have demonstrated a transition to many-body localizing behavior in a universal circuit model composed of Gaussian fermionic evolution and fermionic interaction gates. This behavior is demonstrated by the transition to a logarithmic lightcone, seen clearly in Fig.~\ref{fig:alternatinglightcone} when the disorder is greater than the empirically observed value $\nu_c \sim 0.8$. Though we choose a specific model of alternating interactions and disordered free-fermionic evolution for clarity of presentation, we emphasize that our algorithm is completely general beyond this setting, since any universal quantum circuit can be decomposed as a product of Gaussian fermionic evolution and interaction gates, and does not require an ensemble average in principle. 

For example, it would be interesting to see how the algorithm does to examine the performance of actual near-term quantum algorithms, such as a quantum adiabatic optimization algorithm (QAOA) \cite{farhi2014quantum}, which are characterized by a quantum circuit of the repeating structure seen in Fig.~\ref{fig:alternatingcircuit}, variationally optimized over some parameterization of the repeated unit cell. As our algorithm is naturally suited to such a structure, we therefore expect it to reveal new classes of systems which exhibit localization in this setting as well.

\section*{Acknowledgment}
This work was supported in part by National Science Foundation grants PHY-1521016.

\onecolumngrid

\section*{Appendix A: Modified Cauchy-Binet Formula with Different Background Sets -- Diagrammatic Proof} \label{sec:mcbformuladproof}

Here, we prove Eq.~(\ref{eq:modCBform}) using the diagrammatic proof shown in Fig.~\ref{fig:modCBcircuit}. This figure depicts a particular quantum circuit composed of general Gaussian fermionic evolution on modes $(\vec{B_l}, \overline{B}, \vec{B_r})$, rearrangements between these modes and the ancillary mode-sets $\vec{A}_l$ and $\vec{A}_r$, and a partial trace over the qubits corresponding to the ancillary modes $\vec{A}_r$. As stated in the main text, the equivalence between two different ways of evaluating this circuit implies the identity. This equivalence is given by the operator equality

\begin{align}
&\left(F_{\overline{B} \vec{A}_l} U_{g, 2}\right)^{\dagger} \left\{C_{\vec{S}} \tr_{\vec{A}_r}\left[\left(U_{g, 1} F_{\overline{B} \vec{A}_r}\right)^{\dagger} C_{\vec{\alpha}} \left(U_{g, 1} F_{\overline{B} \vec{A}_r}\right) C_{\vec{S}^{\prime}}^{\dagger} \right] \right\} \left(F_{\overline{B} \vec{A}_l} U_{g, 2} \right) \nonumber \\
&\mbox{\hspace{30mm}}= \tr_{\vec{A}_{r}} \left[\left(U_{g, 1} F_{\overline{B} \vec{A}_r} F_{\overline{B}, \vec{A}_l} U_{g, 2} \right)^{\dagger} C_{\vec{S}} C_{\vec{\alpha}} \left(U_{g, 1} F_{\overline{B} \vec{A}_r} F_{\overline{B} \vec{A}_l} U_{g, 2} \right) C_{\vec{S}^{\prime}}^{\dagger} \right]
\label{eq:opequality}
\end{align}

\noindent where the operators $F_{\overline{B} \vec{A}_r}$ and $F_{\overline{B} \vec{A}_l}$ are rearrangements of fermionic modes. We choose these operators to have corresponding single-particle transition matrices

\begin{align}
\mathbf{f}_{\overline{B} \vec{A}_l} &= \begin{pmatrix}
\mathbf{0} & \mathbf{0} & \mathbf{I} & \mathbf{0} & \mathbf{0} \\
\mathbf{0} & (-1)^{|\vec{S}| + |\vec{S}^{\prime}|} \mathbf{I} & \mathbf{0} & \mathbf{0} & \mathbf{0} \\
\mathbf{I} & \mathbf{0} & \mathbf{0} & \mathbf{0} & \mathbf{0} \\
\mathbf{0} & \mathbf{0} & \mathbf{0} & \mathbf{I} & \mathbf{0} \\
\mathbf{0} & \mathbf{0} & \mathbf{0} & \mathbf{0} & \mathbf{I}
\end{pmatrix} \label{eq:fleft} \\
\mathbf{f}_{\overline{B} \vec{A}_r} &= \begin{pmatrix}
\mathbf{I} & \mathbf{0} & \mathbf{0} & \mathbf{0} & \mathbf{0} \\
\mathbf{0} & \mathbf{I} & \mathbf{0} & \mathbf{0} & \mathbf{0} \\
\mathbf{0} & \mathbf{0} & \mathbf{0} & \mathbf{0} & \mathbf{I} \\
\mathbf{0} & \mathbf{0} & \mathbf{0} & \mathbf{I} & \mathbf{0} \\
\mathbf{0} & \mathbf{0} & \mathbf{I} & \mathbf{0} & \mathbf{0}
\end{pmatrix} \label{eq:fright}
\end{align}

\noindent where $F_{\overline{B} \vec{A}_l}^{\dagger} c_{\mu} F_{\overline{B}, \vec{A}_l} \equiv \left(\mathbf{f}_{\overline{B} \vec{A}_l}\cdot \mathbf{c} \right)_{\mu}$, and similarly for $\mathbf{f}_{\overline{B} \vec{A}_r}$. The blocks in the matrices above act on modes $\vec{A}_l$, $\vec{B_l}$, $\overline{B}$, $\vec{B_r}$, $\vec{A}_r$, respectively. In Eq.~(\ref{eq:opequality}), we made use of the fact that $\vec{S}^{\prime} \subseteq \vec{A}_r$, so $C_{\vec{S}^{\p}}$ commutes with $\left(F_{\overline{B} \vec{A}_l} U_{g, 2}\right)$, which acts as the identity on these modes. The right-hand-side of this equation corresponds to the dot-dashed contraction ordering in Fig.~\ref{fig:modCBcircuit}, and the left-hand-side corresponds to the dotted contraction ordering. Labeling the indices of our block matrices in the same way as in Eq.s~(\ref{eq:fleft}) and (\ref{eq:fright}), we thus have 

\begin{align}
\mathbf{u}_1 \mathbf{f}_{\overline{B} \vec{A}_r} = \begin{bmatrix}
\mathbf{I} & \mathbf{0} & \mathbf{0} & \mathbf{0} & \mathbf{0} \\
\mathbf{0} & \left(\mathbf{u}_1\right)_{\vec{B}_l \vec{B}_l} & \mathbf{0} & \left(\mathbf{u}_1\right)_{\vec{B}_l \vec{B}_r} & \left(\mathbf{u}_1\right)_{\vec{B}_l \overline{B}} \\
\mathbf{0} & \left(\mathbf{u}_1\right)_{\overline{B} \vec{B}_l} & \mathbf{0} & \left(\mathbf{u}_1\right)_{\overline{B} \vec{B}_r} & \left(\mathbf{u}_1\right)_{\overline{B} \overline{B}} \\
\mathbf{0} & \left(\mathbf{u}_1\right)_{\vec{B}_r \vec{B}_l} & \mathbf{0} & \left(\mathbf{u}_1\right)_{\vec{B}_r \vec{B}_r} & \left(\mathbf{u}_1\right)_{\vec{B}_r \overline{B}} \\
\mathbf{0} & \mathbf{0} & \mathbf{I} & \mathbf{0} & \mathbf{0} \end{bmatrix} \label{eq:uf}
\end{align}

\noindent and similarly 

\begin{align}
\mathbf{f}_{\overline{B} \vec{A}_l} \mathbf{u}_2 = \begin{bmatrix}
\mathbf{0} & \left(\widetilde{\mathbf{u}}_2\right)_{\overline{B} \vec{B}_l} & \left(\widetilde{\mathbf{u}}_2\right)_{\overline{B} \overline{B}} & \left(\widetilde{\mathbf{u}}_2\right)_{\overline{B} \vec{B}_r} & \mathbf{0} \\
\mathbf{0} &  \left(\widetilde{\mathbf{u}}_2\right)_{\vec{B}_l \vec{B}_l} & \left(\widetilde{\mathbf{u}}_2\right)_{\vec{B}_l \overline{B}} & \left(\widetilde{\mathbf{u}}_2\right)_{\vec{B}_l \vec{B}_r} & \mathbf{0} \\
\mathbf{I} & \mathbf{0} & \mathbf{0} & \mathbf{0} & \mathbf{0} \\
\mathbf{0} & \left(\widetilde{\mathbf{u}}_2\right)_{\vec{B}_r \vec{B}_l} & \left(\widetilde{\mathbf{u}}_2\right)_{\vec{B}_r \overline{B}} & \left(\widetilde{\mathbf{u}}_2\right)_{\vec{B}_r \vec{B}_r} & \mathbf{0} \\
\mathbf{0} & \mathbf{0} & \mathbf{0} & \mathbf{0} & \mathbf{I} \end{bmatrix} \label{eq:fu} \mathrm{,}
\end{align}

\noindent where $\widetilde{\mathbf{u}}_2$ is as defined in Eq.~(\ref{eq:utdef}) This gives, from the left-hand-side (dot-dashed contraction ordering) of Eq.~(\ref{eq:opequality})

\begin{align}
&\left(F_{\overline{B} \vec{A}_l} U_{g, 2}\right)^{\dagger} \left\{C_{\vec{S}} \tr_{\vec{A}_r}\left[\left(U_{g, 1} F_{\overline{B} \vec{A}_r}\right)^{\dagger} C_{\vec{\alpha}} \left(U_{g, 1} F_{\overline{B} \vec{A}_r}\right) C_{\vec{S}^{\prime}}^{\dagger} \right] \right\} \left(F_{\overline{B} \vec{A}_l} U_{g, 2} \right) \nonumber \\ 
& \mbox{\hspace{20mm}} = 2^{|\vec{A}_r|/2} \sum_{\vb} \det\left[\left(\mathbf{u}_1 \mathbf{f}_{\overline{B} \vec{A}_r} \right)_{\va, \left(\vb, \vec{S}^{\prime} \right)}\right] \left(F_{\overline{B} \vec{A}_l} U_{g, 2} \right)^{\dagger} C_{\vec{S}} C_{\vb} \left(F_{\overline{B} \vec{A}_l} U_{g, 2} \right) \label{eq:leftbubblingline1} \\ 
& \mbox{\hspace{20mm}} = 2^{|\vec{A}_r|/2} \sum_{\vg} \left\{ \sum_{\vb} \det\left[\left(\mathbf{u}_1 \mathbf{f}_{\overline{B} \vec{A}_r} \right)_{\va, \left(\vb, \vec{S}^{\prime} \right)}\right] \det \left[\left(\mathbf{f}_{\overline{B} \vec{A}_l} \mathbf{u}_2 \right)_{\left(\vec{S}, \vb \right), \vg} \right] \right\} C_{\vg} \label{eq:leftbubblingline2} \\
& \mbox{\hspace{20mm}} = 2^{|\vec{A}_r|/2} \sum_{\vg} \left\{\sum_{\vb = \left(\vb_l, \vb_r\right)} (-1)^{\left(|\vec{S}| + |\vec{S}^{\prime}|\right)|\vb_l|}\det\left[\left(\mathbf{u}_1 \right)_{\va, \left(\vb_{l}, \vb_{r}, \vec{S}^{\p} \right)}\right] \det\left[\left(\mathbf{u}_2 \right)_{\left(\vec{S}, \vb_{l}, \vb_r \right), \vg}\right] \right\} C_{\vg} \label{eq:leftbubblingline3} \\ 
& \mbox{\hspace{20mm}} = 2^{|\vec{A}_r|/2} \sum_{\vg} \left\{\sum_{\vb = \left(\vb_l, \vb_r\right)} (-1)^{\left(|\vec{S}| + |\vec{S}^{\prime}|\right)|\vb_l| + |\vec{S}^{\p}| |\vb_r| + |\vec{S}| |\vb_l|}\det\left[\left(\mathbf{u}_1 \right)_{\va, \left(\vb_{l}, \vec{S}^{\p}, \vb_{r} \right)}\right] \det\left[\left(\mathbf{u}_2 \right)_{\left(\vb_{l}, \vec{S}, \vb_r \right), \vg}\right] \right\} C_{\vg} \label{eq:leftbubblingline4} \\
& \mbox{\hspace{20mm}} = 2^{|\vec{A}_r|/2} \sum_{\vg} (-1)^{|\vec{S}^{\p}| (|\vec{S}| - |\vg|)} \left\{\sum_{\vb = \left(\vb_l, \vb_r\right)} \det\left[\left(\mathbf{u}_1 \right)_{\va, \left(\vb_{l}, \vec{S}^{\p}, \vb_{r} \right)}\right] \det\left[\left(\mathbf{u}_2 \right)_{\left(\vb_{l}, \vec{S}, \vb_r \right), \vg}\right] \right\} C_{\vg} \label{eq:leftbubblingline5} \mathrm{.}
\end{align}

\noindent To obtain Eq.~(\ref{eq:leftbubblingline2}), we expanded the Gaussian fermionic evolution of the Majorana configuration by Eq.~(\ref{eq:gfevolution}) (since we take $\vec{A}_r$ to correspond to the fermionic modes on a collection of qubits, $|\vec{A}_r|$ must be even). From Eq.~(\ref{eq:leftbubblingline2}) to Eq.~(\ref{eq:leftbubblingline3}), we used the block-matrix forms in Eq.s~(\ref{eq:uf}) and (\ref{eq:fu}) to re-\"{e}xpress the series in terms of minors of $\mathbf{u}_1$ and $\mathbf{u}_2$ only. To obtain Eq.~(\ref{eq:leftbubblingline4}), we rearranged rows and columns in $\mathbf{u}_1$ and $\mathbf{u}_2$, acquiring a phase. To obtain Eq.~(\ref{eq:leftbubblingline5}), we simplified the phase using the relation $|\vb_l| + |\vb_r| = |\vg| - |\vS|$ (since the sub-matrix of $\mathbf{u}_2$ must be square). Similarly, we have

\begin{align}
\mathbf{u}_1 \mathbf{f}_{\overline{B} \vec{A}_r} \mathbf{f}_{\overline{B} \vec{A}_l} \mathbf{u}_2 = \begin{bmatrix}
\mathbf{0} & \left(\widetilde{\mathbf{u}}_2\right)_{\overline{B} \vec{B}_l} & \left(\widetilde{\mathbf{u}}_2\right)_{\overline{B} \overline{B}} & \left(\widetilde{\mathbf{u}}_2\right)_{\overline{B} \vec{B}_r} & \mathbf{0} \\
\mathbf{0} & \left(\mathbf{u}_1 \right)_{\vec{B}_l \vec{B}} \left(\widetilde{\mathbf{u}}_2 \right)_{\vec{B} \vec{B}_l} &  \left(\mathbf{u}_1 \right)_{\vec{B}_l \vec{B}} \left(\widetilde{\mathbf{u}}_2 \right)_{\vec{B} \overline{B}} &  \left(\mathbf{u}_1 \right)_{\vec{B}_l \vec{B}} \left(\widetilde{\mathbf{u}}_2 \right)_{\vec{B} \vec{B}_r} & \left(\mathbf{u}_1\right)_{\vec{B}_l \overline{B}} \\
\mathbf{0} & \left(\mathbf{u}_1 \right)_{\overline{B} \vec{B}} \left(\widetilde{\mathbf{u}}_2 \right)_{\vec{B} \vec{B}_l} &  \left(\mathbf{u}_1 \right)_{\overline{B} \vec{B}} \left(\widetilde{\mathbf{u}}_2 \right)_{\vec{B} \overline{B}} &  \left(\mathbf{u}_1 \right)_{\overline{B} \vec{B}} \left(\widetilde{\mathbf{u}}_2 \right)_{\vec{B} \vec{B}_r} & \left(\mathbf{u}_1\right)_{\overline{B} \overline{B}} \\
\mathbf{0} & \left(\mathbf{u}_1 \right)_{\vec{B}_r \vec{B}} \left(\widetilde{\mathbf{u}}_2 \right)_{\vec{B} \vec{B}_l} &  \left(\mathbf{u}_1 \right)_{\vec{B}_r \vec{B}} \left(\widetilde{\mathbf{u}}_2 \right)_{\vec{B} \overline{B}} &  \left(\mathbf{u}_1 \right)_{\vec{B}_r \vec{B}} \left(\widetilde{\mathbf{u}}_2 \right)_{\vec{B} \vec{B}_r} & \left(\mathbf{u}_1\right)_{\vec{B}_r \overline{B}} \\
\mathbf{I} & \mathbf{0} & \mathbf{0} & \mathbf{0} & \mathbf{0}
\end{bmatrix} \label{eq:uffu}
\end{align}

\noindent Thus, the right-hand-side (dotted contraction ordering) of Eq.~(\ref{eq:opequality}) gives 

\begin{align}
\tr_{\vec{A}_r}\left[\left(U_{g, 1} F_{\overline{B}\vec{A}_r} F_{\overline{B}\vec{A}_l}U_{g, 2}\right)^{\dagger} C_{\vec{S}} C_{\va} \left(U_{g, 1} F_{\overline{B}\vec{A}_r} F_{\overline{B}\vec{A}_l}U_{g, 2}\right) C_{\vec{S}^{\p}}^{\dagger} \right] &= 2^{|\vec{A}_r|/2}\sum_{\vg} \det\left[\left(\mathbf{u}_1 \mathbf{f}_{\overline{B} \vec{A}_r} \mathbf{f}_{\overline{B} \vec{A}_l} \mathbf{u}_2 \right)_{\left(\vec{S}, \va \right) \left(\vg, \vec{S}^{\p} \right)}\right] C_{\vg} \label{eq:rightbubblingline1} \\
&= 2^{|\vec{A}_r|/2}\sum_{\vg} \det \begin{bmatrix}
\left(\widetilde{\mathbf{u}}_2\right)_{\vec{S} \vg} & \mathbf{0}_{|\vec{S}| \times |\vec{S}^{\p}|} \\ 
\left(\mathbf{u}_1\right)_{\va \vec{B}} \left(\widetilde{\mathbf{u}}_2\right)_{\vec{B} \vg} & \left(\mathbf{u}_1\right)_{\va \vec{S}^{\p}}
\end{bmatrix}
C_{\vg} \label{eq:rightbubblingline2} \\
\tr_{\vec{A}_r}\left[\left(U_{g, 1} F_{\overline{B}\vec{A}_r} F_{\overline{B}\vec{A}_l}U_{g, 2}\right)^{\dagger} C_{\vec{S}} C_{\va} \left(U_{g, 1} F_{\overline{B}\vec{A}_r} F_{\overline{B}\vec{A}_l}U_{g, 2}\right) C_{\vec{S}^{\p}}^{\dagger} \right] &= 2^{|\vec{A}_r|/2} \sum_{\vg} (-1)^{|\vec{S}^{\p}| |\vg|} \det \begin{bmatrix}
\mathbf{0}_{|\vec{S}| \times |\vec{S}^{\p}|} & \left(\mathbf{u}_2\right)_{\vec{S} \vg} \\ 
\left(\mathbf{u}_1\right)_{\va \vec{S}^{\p}} & \left(\mathbf{u}_1\right)_{\va \vec{B}} \left(\widetilde{\mathbf{u}}_2\right)_{\vec{B} \vg}
\end{bmatrix} \mathrm{,} \label{eq:rightbubblingline3}
\end{align}

\noindent where $\vec{B} \equiv (\vec{B}_l, \vec{B}_r)$. From Eq.~(\ref{eq:rightbubblingline1}) to Eq.~(\ref{eq:rightbubblingline2}), we similarly used the block-matrix form of Eq.~(\ref{eq:uffu}) to re-\"{e}xpress the minor in Eq.~(\ref{eq:rightbubblingline1}) in-terms of minors of $\mathbf{u}_1$ and $\widetilde{\mathbf{u}}_2$ only. From Eq.~(\ref{eq:rightbubblingline2}) to Eq.~(\ref{eq:rightbubblingline3}), we again rearranged columns in the matrix determinant, acquiring a phase, and used the fact that $\left(\widetilde{\mathbf{u}}_2\right)_{\vS \vg} = \left(\mathbf{u}_2\right)_{\vS \vg}$, since $\vS$ is disjoint from $\vec{B}_l$.

Setting Eq.s~(\ref{eq:rightbubblingline3}) and (\ref{eq:leftbubblingline5}) equal by Eq.~(\ref{eq:opequality}), canceling corresponding factors of $2^{|\vec{A}_r|/2}$, and using linear independence of the $\{C_{\vg}\}$ gives

\begin{align}
(-1)^{|\vec{S}^{\p}| (|\vec{S}| - |\vg|)} \sum_{\vb = \left(\vb_l, \vb_r\right)} \det\left[\left(\mathbf{u}_1 \right)_{\va, \left(\vb_{l}, \vec{S}^{\p}, \vb_{r} \right)}\right] \det\left[\left(\mathbf{u}_2 \right)_{\left(\vb_{l}, \vec{S}, \vb_r \right), \vg}\right] &= (-1)^{|\vec{S}^{\p}| |\vg|} \det \begin{bmatrix}
\mathbf{0}_{|\vec{S}| \times |\vec{S}^{\p}|} & \left(\mathbf{u}_2\right)_{\vec{S} \vg} \\ 
\left(\mathbf{u}_1\right)_{\va \vec{S}^{\p}} & \left(\mathbf{u}_1\right)_{\va \vec{B}} \left(\widetilde{\mathbf{u}}_2\right)_{\vec{B} \vg}
\end{bmatrix} \\
\sum_{\vb = \left(\vb_l, \vb_r\right)} \det\left[\left(\mathbf{u}_1 \right)_{\va, \left(\vb_{l}, \vec{S}^{\p}, \vb_{r} \right)}\right] \det\left[\left(\mathbf{u}_2 \right)_{\left(\vb_{l}, \vec{S}, \vb_r \right), \vg}\right] &= (-1)^{|\vec{S}^{\p}| |\vec{S}|} \det \begin{bmatrix}
\mathbf{0}_{|\vec{S}| \times |\vec{S}^{\p}|} & \left(\mathbf{u}_2\right)_{\vec{S} \vg} \\ 
\left(\mathbf{u}_1\right)_{\va \vec{S}^{\p}} & \left(\mathbf{u}_1\right)_{\va \vec{B}} \left(\widetilde{\mathbf{u}}_2\right)_{\vec{B} \vg}
\end{bmatrix}
\end{align}

\section*{Appendix B: Modified Cauchy-Binet Formula with Different Background Sets -- Determinental Proof} 
\label{sec:mcbformula}

Here we prove Eq.~(\ref{eq:modCBform}) using determinental identities. Let $\mathbf{u}_1$ and $\mathbf{u}_2$ be orthogonal matrices, and let $\va$, $\vb$, $\vg$, $\vS$, $\vS^{\p}$ be tuples, for which

\begin{align}
|\vb| = |\va| - |\vS^{\p}| = |\vg| - |\vS| \rm{,}
\label{eq:lengthrelation}
\end{align}

\noindent and  

\begin{align}
\begin{cases}
\vS\mathrm{,} \ \vS^{\p} \subseteq \overline{B} & \\
\vec{B} \equiv \left(\vec{B}_l, \vec{B}_r \right) & \\
\vb = \left(\vb_l, \vb_r \right) & \mathrm{where} \ \vb_l \subseteq \vec{B}_l \ \mathrm{and} \ \vb_r \subseteq \vec{B}_r
\end{cases}
\end{align}

\noindent for $\vec{B}$ a contiguous set of indices and $\vec{B}_l$, $\vec{B}_r$, and $\overline{B}$ all disjoint. We will show

\begin{align}
\sum_{\vb} \det{\left[\left(\mathbf{u}_1\right)_{\va \left(\vb_l, \vS^{\p}, \vb_r\right)} \right]} \det{\left[\left(\mathbf{u}_2\right)_{\left(\vb_l, \vS, \vb_r\right) \vg} \right]} = (-1)^{|\vec{S}| |\vec{S}^{\p}|} \det
\begin{bmatrix}
\mathbf{0}_{|\vS| \times |\vS^{\p}|} & \left(\mathbf{u}_2\right)_{\vS \vg} \\
\left(\mathbf{u}_1\right)_{\va \vS^{\p}} & \left(\mathbf{u}_1\right)_{\va \vec{B}} \left(\widetilde{\mathbf{u}}_2\right)_{\vec{B} \vg}
\end{bmatrix} \rm{,}
\end{align}

\noindent where the sum is over all tuples $\vb$ consistent with the constraints. We first rearrange columns in $\mathbf{u}_1$ and $\mathbf{u}_2$ such that the first $\vec{B}$ constitutes the first $|\vec{B}|$ columns, as

\begin{align}
\sum_{\vb} \det{\left[\left(\mathbf{u}_1\right)_{\va \left(\vb_l, \vS^{\p}, \vb_r\right)} \right]} \det{\left[\left(\mathbf{u}_2\right)_{\left(\vb_l, \vS, \vb_r\right) \vg} \right]} = \sum_{\vb} (-1)^{|\vb_l|(|\vS| + |\vS^{\p}|)} \det{\left[\left(\mathbf{u}_1^{\p}\right)_{\va \left(\vS^{\p}, \vb \right)} \right]} \det{\left[\left(\mathbf{u}_2^{\p}\right)_{\left(\vS, \vb \right) \vg} \right]} 
\end{align}

\noindent where $\mathbf{u}_1^{\p}$ and $\mathbf{u}_2^{\p}$ are the rearranged matrices. If $|\vS|$ and $|\vS^{\p}|$ have the same parity, then the sign factor inside the sum is $1$. Otherwise, it evaluates to $(-1)^{|\vb_l|}$, which we absorb onto $\mathbf{u}_2^{\p}$ by multiplying its columns in $\vec{B}_l$ by $(-1)$ to obtain $\widetilde{\mathbf{u}}_2^{\p}$. This gives

\begin{align}
\sum_{\vb} \det{\left[\left(\mathbf{u}_1^{\p}\right)_{\va \left(\vS^{\p}, \vb \right)} \right]} \det{\left[\left(\widetilde{\mathbf{u}}_2^{\p}\right)_{\left(\vS, \vb \right) \vg} \right]} &= \sum_{\vb} \left\{\sum_{\vec{H}} \varepsilon^{\vec{H} \vS^{\p}} \det{\left[\left(\mathbf{u}_1^{\p}\right)_{\vec{H} \vS^{\p}} \right]}\det{\left[\left(\mathbf{u}_1^{\p}\right)_{\va/\vec{H}, \vb } \right]} \right\} \nonumber \\ 
& \mbox{\hspace{20mm}} \times \left\{ \sum_{\vec{L}} \varepsilon^{\vec{L} \vec{S}} \det{\left[\left(\widetilde{\mathbf{u}}_2^{\p}\right)_{\vS \vec{L}} \right]} \det{\left[\left(\widetilde{\mathbf{u}}_2^{\p}\right)_{\vb, \vg/\vec{L}} \right]}\right\} \label{eq:mcbproofline1} \\
&= \sum_{\vec{H}, \vec{L}} \varepsilon^{\vec{H} \vec{S}^{\p}} \varepsilon^{\vec{L} \vec{S}} \det{\left[\left(\mathbf{u}_1^{\p}\right)_{\vec{H} \vS^{\p}} \right]} \det{\left[\left(\widetilde{\mathbf{u}}_2^{\p}\right)_{\vS \vec{L}} \right]} \nonumber \\
&\mbox{\hspace{20mm}} \times \left\{\sum_{\vb}  \det{\left[\left(\mathbf{u}_1^{\p}\right)_{\va/\vec{H}, \vb } \right]} \det{\left[\left(\widetilde{\mathbf{u}}_2^{\p}\right)_{\vb, \vg/\vec{L}} \right]}\right\} \label{eq:mcbproofline2} \\
&= \sum_{\vec{H}, \vec{L}} \varepsilon^{\vec{H} \vS^{\p}} \varepsilon^{\vec{L} \vec{S}} \det{\left[\left(\mathbf{u}_1^{\p}\right)_{\vec{H} \vS^{\p}} \right]} \det{\left[\left(\widetilde{\mathbf{u}}_2^{\p}\right)_{\vS \vec{L}} \right]} \det{\left[\left(\mathbf{u}_1^{\p}\right)_{\va/\vec{H}, \left(\vec{B}_l, \vec{B}_r\right)} \left(\widetilde{\mathbf{u}}_2^{\p}\right)_{\left(\vec{B}_l, \vec{B}_r\right), \vg/\vec{L}}\right]} \label{eq:mcbproofline3} \\
&= \sum_{\vec{L}} \varepsilon^{\vec{L} \vec{S}} \det{\left[\left(\widetilde{\mathbf{u}}_2^{\p}\right)_{\vS \vec{L}} \right]} \det{\begin{bmatrix} \left(\mathbf{u}_1^{\p}\right)_{\va \vS^{\p}} & \left(\mathbf{u}_1^{\p}\right)_{\va, \left(\vec{B}_l, \vec{B}_r\right)} \left(\widetilde{\mathbf{u}}_2^{\p}\right)_{\left(\vec{B}_l, \vec{B}_r\right), \vg/\vec{L}} \end{bmatrix}} \label{eq:mcbproofline4}\\
&= \sum_{\vec{L}} \varepsilon^{\vec{L} \vec{S}} \det{\begin{bmatrix}\mathbf{0}_{|\vec{S}| \times |\vS^{\p}|} & \left(\widetilde{\mathbf{u}}_2^{\p}\right)_{\vS \vec{L}} \end{bmatrix}} \det{\begin{bmatrix} \left(\mathbf{u}_1^{\p}\right)_{\va \vS^{\p}} & \left(\mathbf{u}_1^{\p}\right)_{\va, \left(\vec{B}_l, \vec{B}_r\right)} \left(\widetilde{\mathbf{u}}_2^{\p}\right)_{\left(\vec{B}_l, \vec{B}_r\right), \vg/\vec{L}} \end{bmatrix}} \label{eq:mcbproofline5} \\
\sum_{\vb} \det{\left[\left(\mathbf{u}_1^{\p}\right)_{\va \left(\vS, \vb \right)} \right]} \det{\left[\left(\widetilde{\mathbf{u}}_2^{\p}\right)_{\left(\vS^{\p}, \vb \right) \vg} \right]} &= (-1)^{|\vS^{\p}| |\vec{S}|} \det
\begin{bmatrix}
\mathbf{0}_{|\vS| \times |\vS^{\p}|} & \left(\widetilde{\mathbf{u}}_2^{\p}\right)_{\vS \vg} \\
\left(\mathbf{u}_1^{\p}\right)_{\va \vS^{\p}} & \left(\mathbf{u}_1^{\p}\right)_{\va \left(\vec{B}_l, \vec{B}_r\right)} \left(\widetilde{\mathbf{u}}_2^{\p}\right)_{\left(\vec{B}_l, \vec{B}_r\right) \vg} \label{eq:mcbproofline6}
\end{bmatrix}
\end{align} 

\noindent From Eq.~(\ref{eq:mcbproofline1}) to Eq.~(\ref{eq:mcbproofline2}), we used the Laplace expansion by complementary minors formula, where $\varepsilon^{\vec{H} \vec{L}} = (-1)^{\sum_{i = 1}^{|\vec{H}|}H_i + \sum_{i = 1}^{|\vec{L}|}L_i}$. From Eq.~(\ref{eq:mcbproofline2}) to Eq.~(\ref{eq:mcbproofline3}), we used the Cauchy-Binet formula for the sum on $\vb$. From Eq.~(\ref{eq:mcbproofline3}) to Eq.~(\ref{eq:mcbproofline4}), we identify the sum on $\vec{H}$ with the Laplace expansion by complementary minors. From Eq.~(\ref{eq:mcbproofline4}) to Eq.~(\ref{eq:mcbproofline5}), we include a block of zeroes in $\widetilde{\mathbf{u}}_2^{\p}$ so as to identify the sum on $\vec{L}$ as a second Laplace expansion by complementary minors from Eq.~(\ref{eq:mcbproofline5}) to Eq.~(\ref{eq:mcbproofline6}). However, since including this block of zeroes shifts the indices of $\vec{L}$ by $|\vec{S}^{\p}|$, this incurs an additional factor of $(-1)^{|\vS||\vS^{\p}|}$ from $\varepsilon^{\vec{L} \vec{S}}$. 

Finally, we may rearrange columns and use the fact that $\left(\widetilde{\mathbf{u}}_2\right)_{\vec{S}\vg} = \left(\mathbf{u}_2\right)_{\vec{S}\vg}$ to recover the formula

\begin{align}
\sum_{\vb} \det{\left[\left(\mathbf{u}_1\right)_{\va \left(\vb_l, \vS, \vb_r\right)} \right]} \det{\left[\left(\mathbf{u}_2\right)_{\left(\vb_l, \vS^{\p}, \vb_r\right) \vg} \right]} = (-1)^{|\vec{S}| |\vec{S}^{\p}|} \det
\begin{bmatrix}
\mathbf{0}_{|\vS^{\p}| \times |\vS|} & \left(\mathbf{u}_2\right)_{\vS^{\p} \vg} \\
\left(\mathbf{u}_1\right)_{\va \vS} & \left(\mathbf{u}_1\right)_{\va \vec{B}} \left(\widetilde{\mathbf{u}}_2\right)_{\vec{B} \vg}
\end{bmatrix} \rm{.}
\label{eq:MCBformuladiffS}
\end{align}

\section*{Appendix C: Exact Formula for the OTO Commutator}
\label{sec:exactoto}

Let a unitary consisting of one interaction gate with two periods of Gaussian fermionic be given by $U = U_{g, 1} V_j U_{g, 2}$, for Gaussian fermionic operations $U_{g, \{1, 2\}}$ and $V_j = \exp\left(-\frac{i\pi}{4}Z_j Z_{j + 1}\right) \equiv V$. As in the main text, let $\vec{q}_j = (2j - 1, 2j, 2j + 1, 2(j + 1)) \equiv \vec{q}$ (since the index $j$ can be seen from context). From Eq.~(\ref{eq:cfrelation}), we see that it suffices to calculate

\begin{align}
d^{-1} \tr\left[AB(t)AB(t)\right] = d^{-1} (-1)^{\frac{1}{2} \left[|\va| (|\va| - 1) + |\ve| (|\ve| - 1)\right]} \tr\left[C_{\ve} \left( U_{g, 1} V_j U_{g, 2} \right)^{\dagger} C_{\va} \left( U_{g, 1} V_j U_{g, 2} \right) C_{\ve} \left( U_{g, 1} V_j U_{g, 2} \right)^{\dagger} C_{\va} \left( U_{g, 1} V_j U_{g, 2} \right) \right] \mathrm{,}
\end{align}

\noindent where $A = i^a C_{\ve}$ and $B = i^b C_{\va}$ for integers $a$ and $b$, and for which $i^{2a} = (-1)^{\frac{1}{2}|\va|(|\va| - 1)}$ and $i^{2b} = (-1)^{\frac{1}{2}|\ve|(|\ve| - 1)}$ from the relations $A^2 = B^2 = I$ (we have assumed $A$ and $B$ to be Hermitian and unitary). From Eq.~(\ref{eq:gfevolution}), we have

\begin{align}
U_{g, 1}^{\dagger} C_{\va} U_{g, 1} = \sum_{\vb} \det\left[\left(\mathbf{u}_1\right)_{\va \vb}\right] C_{\vb}
\end{align}

\noindent Let $\vb = (\vb_{q} \cup \vb_{\overline{q}})$, where $\vb_{q} = \vb \cap \vec{q}$ and $\vb_{\overline{q}} = \vb \cap \overline{q}$, for $\overline{q}$ the complement of $\vec{q}$ in $[2n] \equiv (1, 2, \dots, 2n)$, the full set of modes. Since $\vec{q}$ is contiguous, we can apply Eq.~(\ref{eq:MCBformuladiffS}) to obtain

\begin{align}
\left(U_{g, 1} V U_{g, 2}\right)^{\dagger} C_{\va} \left(U_{g, 1} V U_{g, 2}\right) &= \sum_{\vb} \det\left[\left(\mathbf{u}_1\right)_{\va \vb}\right] \left(V U_{g, 2}\right)^{\dagger} C_{\vb} \left(V U_{g, 2}\right) \label{eq:iterproofline1} \\
&= \sum_{k = 0}^{\min(|\vec{q}|, |\va|)} \sum_{\{\vb_q| |\vb_q| = k\}}\sum_{\vb_{\overline{q}}} \det\left[\left(\mathbf{u}_1\right)_{\va, \left(\vb_q \cup \vb_{\overline{q}} \right)}\right] \left(V U_{g, 2}\right)^{\dagger} C_{ \left(\vb_q \cup \vb_{\overline{q}} \right)}\left(V U_{g, 2}\right) \label{eq:iterproofline2} \\
&= \sum_{k = 0}^{\min(|\vec{q}|, |\va|)} \sum_{\{\vb_q| |\vb_q| = k\}} i^{\varphi(\vb_{q})}\sum_{\vb_{\overline{q}}} \det\left[\left(\mathbf{u}_1\right)_{\va, \left(\vb_q \cup \vb_{\overline{q}} \right)}\right] U_{g, 2}^{\dagger} C_{ \left(\mathcal{V}(\vb_q) \cup \vb_{\overline{q}} \right)} U_{g, 2} \label{eq:iterproofline3} \\
&= \sum_{k = 0}^{\min(|\vec{q}|, |\va|)} \sum_{\{\vb_q| |\vb_q| = k\}} i^{\varphi(\vb_{q})}\sum_{\vb_{\overline{q}}} \det\left[\left(\mathbf{u}_1\right)_{\va, \left(\vb_q \cup \vb_{\overline{q}} \right)}\right] \sum_{\vg} \det\left[\left(\mathbf{u}_2\right)_{\left(\mathcal{V}(\vb_q) \cup \vb_{\overline{q}}\right), \vg}\right] C_{\vg} \label{eq:iterproofline4} \\
&= \sum_{k = 0}^{\min(|\vec{q}|, |\va|)} \sum_{\{\vb_q| |\vb_q| = k\}} i^{\varphi(\vb_{q})}\sum_{\vg} \left\{\sum_{\vb_{\overline{q}}} \det\left[\left(\mathbf{u}_1\right)_{\va, \left(\vb_q \cup \vb_{\overline{q}} \right)}\right] \det\left[\left(\mathbf{u}_2\right)_{\left(\mathcal{V}(\vb_q) \cup \vb_{\overline{q}}\right), \vg}\right]\right\} C_{\vg} \label{eq:iterproofline5} \\
&= \sum_{\vg} \left[ \sum_{\{\left(k, \vb_q\right)| |\vb_q| = k\}} i^{\varphi(\vb_{q})} (-1)^{|\vb_q| |\mathcal{V}(\vb_q)|} \det
\begin{bmatrix}
\mathbf{0}_{|\mathcal{V}(\vb_q)| \times |\vb_q|} & \left(\mathbf{u}_2\right)_{\mathcal{V}(\vb_q) \vg} \\
\left(\mathbf{u}_1\right)_{\va \vb_q} & \left(\mathbf{u}_1\right)_{\va \overline{q}} \left(\mathbf{u}_2\right)_{\overline{q} \vg}
\end{bmatrix}\right] C_{\vg}\label{eq:iterproofline6} \\
\left(U_{g, 1} V U_{g, 2}\right)^{\dagger} C_{\va} \left(U_{g, 1} V U_{g, 2}\right) &\equiv  \sum_{\vg} \left( \sum_{\{\left(k, \vb_q\right)| |\vb_q| = k\}} i^{\varphi(\vb_{q})} (-1)^{|\vb_q| |\mathcal{V}(\vb_q)|} \det \left\{\left[\mathbf{u}_{(1, 2)}\right]_{\left[\mathcal{V}(\vb_q), \va \right], \left[\vb_q, \vg \right]}\right\} \right) C_{\vg} \label{eq:iterproofline7} \mathrm{.}
\end{align}

\noindent In Eq.~(\ref{eq:iterproofline2}), we split the sum into sums over $\vb_q$, $\vb_{\overline{q}}$. In  Eq.~(\ref{eq:iterproofline3}), we let

\begin{align}
V^{\dagger}C_{\vb_q} V = i^{\varphi \left(\vb_q \right)} C_{\mathcal{V}(\vb_q)} \mathrm{.}
\end{align}

\noindent using the fact that $V$ commutes with any modes not in $\vec{q}$, and that the set $\vec{q}$ is contiguous. From Eq.s~(\ref{eq:iterproofline3})-(\ref{eq:iterproofline6}), we applied Eq.~(\ref{eq:MCBformuladiffS}) (ang grouped sums for notational convenience). In Eq.~(\ref{eq:iterproofline7}), we defined

\begin{align}
\mathbf{u}_{(1, 2)} \equiv \begin{bmatrix}
\mathbf{0}_{|\vec{q}| \times |\vec{q}|} & \left(\mathbf{u}_2\right)_{\vec{q} [2n]} \\
\left(\mathbf{u}_1\right)_{[2n] \vec{q}} & \left(\mathbf{u}_1\right)_{[2n] \overline{q}} \left(\mathbf{u}_2\right)_{\overline{q} [2n]}
\end{bmatrix} \mathbf{.}
\end{align}

\noindent It is straightforward to show that $\mathbf{u}_{(1, 2)}$ is itself orthogonal for orthogonal $\mathbf{u}_{\{1, 2\}}$, as

\begin{align}
\mathbf{u}_{(1, 2)} \mathbf{u}_{(1, 2)}^{\mathrm{T}} &= \begin{bmatrix}
\mathbf{0}_{|\vec{q}| \times |\vec{q}|} & \left(\mathbf{u}_2\right)_{\vec{q} [2n]} \\
\left(\mathbf{u}_1\right)_{[2n] \vec{q}} & \left(\mathbf{u}_1\right)_{[2n] \overline{q}} \left(\mathbf{u}_2\right)_{\overline{q} [2n]}
\end{bmatrix} \begin{bmatrix}
\mathbf{0}_{|\vec{q}| \times |\vec{q}|} & \left(\mathbf{u}^{\mathrm{T}}_1\right)_{\vec{q}[2n]} \\
\left(\mathbf{u}^{\mathrm{T}}_2\right)_{[2n] \vec{q}} & \left(\mathbf{u}^{\mathrm{T}}_2\right)_{[2n] \overline{q}} \left(\mathbf{u}^{\mathrm{T}}_1\right)_{\overline{q} [2n]}
\end{bmatrix} \label{eq:orthoproofline1} \\
&= \begin{bmatrix}
\left(\mathbf{u}_2\right)_{\vec{q} [2n]} \left(\mathbf{u}^{\mathrm{T}}_2\right)_{[2n] \vec{q}} & \left(\mathbf{u}_2\right)_{\vec{q} [2n]} \left(\mathbf{u}_2^\mathrm{T} \right)_{[2n] \overline{q}} \left(\mathbf{u}_1^{\mathrm{T}}\right)_{\overline{q} [2n]} \\
\left(\mathbf{u}_1\right)_{[2n] \overline{q}} \left(\mathbf{u}_2\right)_{\overline{q} [2n]} \left(\mathbf{u}_2^{\mathrm{T}}\right)_{[2n] \vec{q}} & \left(\mathbf{u}_1\right)_{[2n] \vec{q}} \left(\mathbf{u}^{\mathrm{T}}_1\right)_{\vec{q} [2n]} +  \left(\mathbf{u}_1\right)_{[2n] \overline{q}} \left(\mathbf{u}_2\right)_{\overline{q} [2n]}  \left(\mathbf{u}^{\mathrm{T}}_2\right)_{[2n] \overline{q}} \left(\mathbf{u}^{\mathrm{T}}_1\right)_{\overline{q} [2n]} \\
\end{bmatrix} \label{eq:orthoproofline2} \\
&= \begin{bmatrix}
\mathbf{I}_{\vec{q} \vec{q}} & \mathbf{I}_{\vec{q} \overline{q}} \left(\mathbf{u}^{\mathrm{T}}_1\right)_{\overline{q} [2n]} \\
\left(\mathbf{u}_1\right)_{[2n] \overline{q}} \mathbf{I}_{\overline{q} \vec{q}} & \left(\mathbf{u}_1\right)_{[2n] \vec{q}} \left(\mathbf{u}^{\mathrm{T}}_1\right)_{\vec{q} [2n]} + \left(\mathbf{u}_1\right)_{[2n] \overline{q}} \left(\mathbf{u}_1^{\mathrm{T}}\right)_{\overline{q}[2n]}
\end{bmatrix} \label{eq:orthoproofline3} \\
&= \begin{pmatrix}
\mathbf{I}_{\vec{q} \vec{q}} & \mathbf{0}_{\vec{q} [2n]} \\
\mathbf{0}_{[2n] \vec{q}} & \mathbf{I}_{[2n][2n]}
\end{pmatrix} \label{eq:orthoproofline4} \\
\mathbf{u}_{(1, 2)} \mathbf{u}_{(1, 2)}^{\mathrm{T}} &= \mathbf{I} \label{eq:orthoproofline5}
\end{align}

\noindent In Eq.~(\ref{eq:orthoproofline2}), we applied the orthogonality of $\mathbf{u}_{\{1, 2\}}$ when contracted along the indices $[2n]$. From Eq.~(\ref{eq:orthoproofline3}) to Eq.~(\ref{eq:orthoproofline4}), we used the facts

\begin{align}
\begin{cases}
\mathbf{I}_{\vec{q} \overline{q}} = \mathbf{0}_{\vec{q} \overline{q}} \\ 
\left(\mathbf{u}_1\right)_{[2n] \vec{q}} \left(\mathbf{u}^{\mathrm{T}}_1\right)_{\vec{q} [2n]} + \left(\mathbf{u}_1\right)_{[2n] \overline{q}} \left(\mathbf{u}_1^{\mathrm{T}}\right)_{\overline{q}[2n]} = \mathbf{u}_1 \mathbf{u}^{\mathrm{T}}_1 = \mathbf{I}
\end{cases}
\end{align}

\noindent It is straightforward to show that $\mathbf{u}_{(1, 2)}^{\mathrm{T}} \mathbf{u}_{(1, 2)} = \mathbf{I}$ as well. We can therefore iterate this procedure for conjugation by an additional interaction gate, $V^{\p}$, acting on the subset of qubits $\vec{q}^{\p}$, as

\begin{align}
U^{\dg} C_{\va} U &\equiv \left(U_{g, 1} V U_{g, 2} V^{\p} U_{g, 3}\right)^{\dg} C_{\va} \left(U_{g, 1} V U_{g, 2} V^{\p} U_{g, 3}\right) \label{eq:seconditerline1} \\
&= \sum_{\vl} \left[\sum_{\substack{\left\{\left(k, \vb_q \right) | |\vb_q| = k \right\} \\ \left\{\left(k^{\p}, \vg_{q^{\p}} \right) | |\vg_{q^{\p}}| = k^{\p} \right\}}} i^{\varphi\left(\vb_q\right) + \varphi\left(\vg_{q^{\p}}\right)} (-1)^{|\vb_q| |\mathcal{V}(\vb_q)|} \right. \label{eq:seconditerline2} \\ 
& \mbox{\hspace{30mm}} \times \left. \left(\sum_{\vg} \det \left\{\left[\mathbf{u}_{\left(1, 2\right)}\right]_{\left[\mathcal{V}\left(\vb_q \right), \va \right] \left[\vb_q, \left(\vg \cup \vg_{q^{\p}} \right)\right]}\right\} \det \left[\left(\mathbf{u}_3\right)_{\left[\vg \cup \mathcal{V}\left(\vg_{q^{\p}}\right)\right] \vl}\right] \right) \right] C_{\vl} \nonumber \\
&= \sum_{\vl} \left\{\sum_{\substack{\left\{\left(k, \vb_q \right) | |\vb_q| = k \right\} \\ \left\{\left(k^{\p}, \vg_{q^{\p}} \right) | |\vg_{q^{\p}}| = k^{\p} \right\}}} i^{\varphi\left(\vb_q\right) + \varphi\left(\vg_{q^{\p}}\right)} (-1)^{|\vb_q| |\mathcal{V}\left(\vb_q\right)|} \right. \label{eq:seconditerline3} \\ 
&\mbox{\hspace{30mm}}\times \left. (-1)^{\left(|\vb_q| + |\vg_{q^{\p}}|\right)|\mathcal{V}\left(\vg_{q^{\p}} \right)|}\det \begin{pmatrix} 
\mathbf{0}_{|\mathcal{V}\left(\vg_{q^{\p}}\right)| \times \left(|\vb_q| + |\vg_{q^{\p}}|\right)} & \left(\mathbf{u}_3 \right)_{\mathcal{V}\left(\vg_{q^{\p}} \right) \vl} \\
\left[\mathbf{u}_{\left(1, 2\right)}\right]_{\left[\mathcal{V}(\vb_q), \va \right]\left(\vb_q, \vg_{q^{\p}}\right)} & \left[\mathbf{u}_{\left(1, 2\right)}\right]_{\left[\mathcal{V}\left(\vb_q\right), \va \right]\overline{q}^{\p}} \left(\mathbf{u}_3\right)_{\overline{q}^{\p} \vl}
\end{pmatrix}\right\} C_{\vl} \nonumber \\ 
U^{\dg} C_{\va} U &= \sum_{\vl} \left(\sum_{\substack{\left\{\left(k, \vb_q \right) | |\vb_q| = k \right\} \\ \left\{\left(k^{\p}, \vg_{q^{\p}} \right) | |\vg_{q^{\p}}| = k^{\p} \right\}}} i^{\varphi\left(\vb_q\right) + \varphi\left(\vg_{q^{\p}}\right)} (-1)^{|\vb_q| |\mathcal{V}\left(\vb_q\right)| + |\vg_{q^{\p}}| |\mathcal{V}\left(\vg_{q^{\p}}\right)|} \det\left\{\left[\mathbf{u}_{(1, 2, 3)}\right]_{\left[\mathcal{V}\left(\vg_{q^{\p}}\right), \mathcal{V}\left(\vb_{q}\right), \va \right]\left[\vg_{q^{\p}}, \vb_q, \vl \right]}\right\} \right) C_{\vl} \label{eq:seconditerline4} \mathrm{,}
\end{align}

\noindent where

\begin{align}
\mathbf{u}_{\left(1, 2, 3\right)} \equiv 
\begin{bmatrix}
\mathbf{0}_{|\vec{q}^{\p}| \times |\vec{q}^{\p}|} & \mathbf{0}_{|\vec{q}^{\p}| \times |\vec{q}|} & \left(\mathbf{u}_3\right)_{\vec{q}^{\p} [2n]} \\
\left(\mathbf{u}_2 \right)_{\vec{q} \vec{q}^{\p}} & \mathbf{0}_{|\vec{q}| \times |\vec{q}|} & \left(\mathbf{u}_{2}\right)_{\vec{q} \overline{q}^{\prime}} \left(\mathbf{u}_{3}\right)_{\overline{q}^{\prime} [2n]} \\
\left(\mathbf{u}_1\right)_{[2n] \overline{q}} \left(\mathbf{u}_2\right)_{\overline{q} \vec{q}} & \left(\mathbf{u}_1\right)_{[2n] \vec{q}} & \left(\mathbf{u}_1\right)_{[2n] \overline{q}} \left(\mathbf{u}_2\right)_{\overline{q}\overline{q}^{\p}} \left(\mathbf{u}_3\right)_{\overline{q}^{\p} [2n]} \rm{.}
\end{bmatrix}
\end{align}

\noindent From Eq.~(\ref{eq:seconditerline3}) to Eq.~(\ref{eq:seconditerline4}), we cancelled a phase of $(-1)^{|\vb_{q}||\mathcal{V}(\vg_{q^{\p}})|}$ by exchanging columns $\vb_q \leftrightarrow \vg_{{q}^{\p}}$ inside the determinant to yield a phase of $(-1)^{|\vb_q| |\vg_{{q}^{\p}}|}$ and used the fact that $(-1)^{|\vb_q|\left(|\vg_{q^{\p}}| + |\mathcal{V}\left(\vg_{q^{\p}}\right)|\right)} = 1$ by the parity-preserving property of $V$. It is clear that $\mathbf{u}_{\left(1, 2, 3\right)}$ is orthogonal by the orthogonality property of $\mathbf{u}_{\left(1, 2\right)}$. We can continue to iterate this process, $g$ times for $g$ interaction gates present, incurring an ancillary set of modes $\vec{q}_i$ for every iteration $i \in \{1, 2, \dots, g\}$. Let $\vec{B} = \bigcup_{i = 1}^g \vb_i$ for $\vb_i \subseteq \vec{q}_i$, and let $\mathbf{M}$ be the orthogonal matrix obtained as the result of these iterations. We have
	
\begin{align}
& \left(U_{g, 1} V_1 U_{g, 2} \dots V_n U_{g, n + 1}\right)^{\dg} C_{\va} \left(U_{g, 1} V_1 U_{g, 2} \dots V_n U_{g, n + 1}\right) =  \sum_{\vec{B} \subseteq \bigcup_{i = 1}^g \vec{q}_i} i^{\sum_{i = 1}^g \varphi\left(\vb_i \right)} (-1)^{\sum_{i = 1}^g |\vb_i| |\mathcal{V}\left(\vb_i \right)|} \nonumber \\ 
& \mbox{\hspace{90mm}} \times \left(\sum_{\vg} \det \left\{\mathbf{M}_{\left[\mathcal{V}^{\times g} \left(\vec{B}\right), \va \right] \left[\vec{B}, \vg \right]}\right\} \right) C_{\vg} \mathrm{,}
\end{align}

\noindent where the $\vec{\beta}_i \subseteq \vec{B}$ are ordered in descending order of $i$ when indexing the rows and columns of $\mathbf{M}$ inside the determinant, and the length $|\vg|$ of the tuple $\vg$ is such that the determinant inside the matrix is square.

We next calculate the OTO correlator as

\begin{align}
&\tr\left[C_{\ve} \left(U^{\dg} C_{\va} U\right) C_{\ve} \left(U^{\dg} C_{\va} U\right)\right] = \sum_{\vec{B}, \vec{B}^{\p} \subseteq \bigcup_{i = 1}^g \vec{q}_i} i^{\sum_{i = 1}^g \left[\varphi\left(\vb_i \right) + \varphi\left(\vb^{\p}_i \right) \right]} \label{eq:otocexactline1} \\
 & \mbox{\hspace{10mm}} \times (-1)^{\sum_{i = 1}^g \left[|\vb_i| |\mathcal{V}\left(\vb_i \right)| + |\vb^{\p}_i| |\mathcal{V}\left(\vb^{\p}_i \right)|\right]} \sum_{\vg, \vg^{\p}} \det \left\{\mathbf{M}_{\left[\mathcal{V}^{\times g} \left(\vec{B}\right), \va \right] \left[\vec{B}, \vg \right]}\right\} \det \left\{\mathbf{M}_{\left[\mathcal{V}^{\times g} \left(\vec{B}^{\p}\right), \va \right] \left[\vec{B}^{\p}, \vg^{\p} \right]}\right\} \tr\left(C_{\ve} C_{\vg} C_{\ve} C_{\vg^{\p}}\right) \nonumber \\
& \frac{1}{2^n}(-1)^{\frac{1}{2}|\ve|(\ve - 1)}\tr\left[C_{\ve} \left(U^{\dg} C_{\va} U\right) C_{\ve} \left(U^{\dg} C_{\va} U\right)\right] = \sum_{\vec{B}, \vec{B}^{\p} \subseteq \bigcup_{i = 1}^g \vec{q}_i} i^{\sum_{i = 1}^g \left[\varphi\left(\vb_i \right) + \varphi\left(\vb^{\p}_i \right) \right]} \label{eq:otocexactline2} \\
& \mbox{\hspace{10mm}} \times (-1)^{\sum_{i = 1}^g \left[ |\vb_i| |\mathcal{V}\left(\vb_i \right)| + |\vb^{\p}_i| |\mathcal{V}\left(\vb^{\p}_i \right)| \right]} \sum_{\vg, \vg^{\p}} (-1)^{\frac{1}{2}|\vg|(|\vg| - 1) + |\ve||\vg| + |\ve \cap \vg|} \det \left\{\mathbf{M}_{\left[\mathcal{V}^{\times g} \left(\vec{B}\right), \va \right] \left[\vec{B}, \vg \right]}\right\} \det \left\{\mathbf{M}_{\left[\mathcal{V}^{\times g} \left(\vec{B}^{\p}\right), \va \right] \left[\vec{B}^{\p}, \vg^{\p} \right]}\right\} \delta_{\vg \vg^{\p}} \nonumber \\
& \frac{1}{2^n}(-1)^{\frac{1}{2}\left[|\ve|(\ve - 1) + |\va|(\va - 1)\right]}\tr\left[C_{\ve} \left(U^{\dg} C_{\va} U\right) C_{\ve} \left(U^{\dg} C_{\va} U\right)\right] = \sum_{\vec{B}, \vec{B}^{\p} \subseteq \bigcup_{i = 1}^g \vec{q}_i} i^{\sum_{i = 1}^g \left[\varphi\left(\vb_i \right) + \varphi\left(\vb^{\p}_i \right)\right]} \label{eq:otocexactline3} \\
& \mbox{\hspace{10 mm}} \times (-1)^{|\vec{B}| + |\vec{B}^{\p}| + \frac{1}{2}\sum_{i = 1}^g \left[|\mathcal{V}\left(\vb_i \right)| - |\vb_i|\right]} \sum_{\vg, \vg^{\p}} \det \left\{\mathbf{M}_{\left[\mathcal{V}^{\times g} \left(\vec{B}\right), \va \right] \left[\vec{B}, \vg \right]}\right\} \det\left[(-1)^{|\ve|} (\mathbf{I} - 2\mathbf{P}_{\ve})_{\vg \vg^{\p}}\right] \det \left\{\mathbf{M}_{\left[\mathcal{V}^{\times g} \left(\vec{B}^{\p}\right), \va \right] \left[\vec{B}^{\p}, \vg^{\p} \right]}\right\} \nonumber \\
&\frac{1}{2^n}(-1)^{\frac{1}{2}\left[|\ve|(\ve - 1) + |\va|(\va - 1)\right]}\tr\left[C_{\ve} \left(U^{\dg} C_{\va} U\right) C_{\ve} \left(U^{\dg} C_{\va} U\right)\right] = \sum_{\vec{B}, \vec{B}^{\p} \subseteq \bigcup_{i = 1}^g \vec{q}_i} i^{\sum_{i = 1}^g \left[\varphi\left(\vb_i \right) + \varphi\left(\vb^{\p}_i \right)\right]} \label{eq:otocexactline4} \\
& \mbox{\hspace{10 mm}} \times (-1)^{|\vec{B}| + |\vec{B}^{\p}| + |\vec{B}||\vec{B}^{\p}| + \frac{1}{2}\sum_{i = 1}^g \left[|\mathcal{V}\left(\vb_i \right)| - |\vb_i|\right]} \det \begin{pmatrix}
\mathbf{0}_{|\vec{B}^{\p}| \times |\vec{B}|} & \mathbf{M}^{\mathrm{T}}_{\vec{B}^{\p} \left[\mathcal{V}^{\times g} \left(\vec{B}^{\p}\right), \va \right]} \\
\mathbf{M}_{\left[\mathcal{V}^{\times g}\left(\vec{B}\right), \va\right] \vec{B}} & 
(-1)^{|\ve|}\mathbf{M}_{\left[\mathcal{V}^{\times g}\left(\vec{B}\right), \va\right] [2n]} \left(\mathbf{I} - 2 \mathbf{P}_{\ve}\right) \mathbf{M}^{\mathrm{T}}_{[2n] \left[\mathcal{V}^{\times g} \left(\vec{B}^{\p}\right), \va \right]} 
\end{pmatrix}  \label{eq:otocexactline5} \\
&\frac{1}{2^n}(-1)^{\frac{1}{2}\left[|\ve|(\ve - 1) + |\va|(\va - 1)\right]}\tr\left[C_{\ve} \left(U^{\dg} C_{\va} U\right) C_{\ve} \left(U^{\dg} C_{\va} U\right)\right] = \sum_{\vec{B}, \vec{B}^{\p} \subseteq \bigcup_{i = 1}^g \vec{q}_i} (-1)^{|\vec{B}| + |\vec{B}^{\p}| + |\vec{B}| |\vec{B}^{\p}|} \nonumber \\ 
& \mbox{\hspace{10mm}} \times (-1)^{\sum_{i = 1}^{g} \left[ |\vb_i| \left(\sum_{j \in \vb_i} j \right) + |\vb^{\p}_i| \left(\sum_{j \in \vb_i^{\p}} j \right) + \delta_{|\vb_i|, 3} + \delta_{|\vb^{\p}_i|, 3} \right]} \det\left[ \mathbf{K}\left(\ve\right)_{\left[\vec{B}^{\p}, \mathcal{V}^{\times g} \left(\vec{B}\right), \va \right] \left[\vec{B}, \mathcal{V}^{\times g} \left(\vec{B}^{\p} \right), \va \right]}\right] \label{eq:otocexactline6} \mathrm{,}
\end{align}

\noindent where, letting $\vec{Q} = \bigcup_{i = 1}^g \vec{q}_i$,

\begin{align}
\mathbf{K}\left(\ve\right) = \begin{pmatrix}
\mathbf{0}_{|\vec{Q}| \times |\vec{Q}|} & \mathbf{M}^{\mathrm{T}}_{\vec{Q} \left[\vec{Q}, [2n]\right]} \\
\mathbf{M}_{\left[\vec{Q}, [2n]\right] \vec{Q}} & (-1)^{|\ve|} \mathbf{M}_{\left[\vec{Q}, [2n]\right] [2n]} \left(\mathbf{I} - 2\mathbf{P}_{\ve} \right) \mathbf{M}^{\mathrm{T}}_{[2n] \left[\vec{Q}, [2n]\right]}
\end{pmatrix} \mathrm{.}
\end{align}

\noindent From Eq.~(\ref{eq:otocexactline2}) to Eq.~(\ref{eq:otocexactline3}), we use the fact that

\begin{align}
(-1)^{\frac{1}{2}\left(|\va| + 2k \right)(|\va| + 2k - 1)} &= (-1)^{\frac{1}{2}|\va|\left(|\va| - 1\right) + k \left(2|\va| - 1\right) + 2k^2} \\
(-1)^{\frac{1}{2}\left(|\va| + 2k \right)(|\va| + 2k - 1)} &= (-1)^{\frac{1}{2}|\va|\left(|\va| - 1\right)} (-1)^k
\end{align}

\noindent for $|\vg| = |\va| + 2k$ and $2k = |\mathcal{V}\left(\vec{B}\right)| - |\vec{B}|$. The latter quantity is guaranteed to be even by the parity-preserving property of $\mathcal{V}$. By the same property, we have

\begin{align}
(-1)^{\sum_{i = 1}^g \left(|\vb_i| |\mathcal{V}\left(\vb_i\right)| + |\vb^{\p}_i| |\mathcal{V}\left(\vb^{\p}_i\right)|\right)} &= (-1)^{\sum_{i = 1}^g \left(|\vb_i| + |\vb^{\p}_i|\right)} \\
(-1)^{\sum_{i = 1}^g \left(|\vb_i| |\mathcal{V}\left(\vb_i\right)| + |\vb^{\p}_i| |\mathcal{V}\left(\vb^{\p}_i\right)|\right)} &= (-1)^{|\vec{B}|  + |\vec{B}^{\p}|}
\end{align}

\noindent From Eq.~(\ref{eq:otocexactline5}) to Eq.~(\ref{eq:otocexactline6}), we used the particular form for the function $\varphi(\vb_j)$, from

\begin{align}
V_i^{\dg} C_{\vb_i} V_i =
\begin{cases}
C_{\vb_i} & |\vb_i| \ \mathrm{even} \\
(-i)(-1)^{\sum_{j \in \vb_i} j} C_{\overline{\beta}_i} & |\vb_i| \ \mathrm{odd}
\end{cases} \label{eq:interphase} \mathrm{,}
\end{align}

\noindent where $\overline{\beta}_i = \vec{q}_i/\vb_i$, and the phase comes from the fact that $Z_j Z_{j + 1} = -C_{\vec{q}_j}$ and $C_{\vec{q}_j} c_{k} = (-1)^{|\vec{q}_j| - k} C_{\vec{q}_j/k}$ for $|\vec{q}_j| = 4$. We see there is a factor of $-i$ for every $\vb_i$ for which $|\vb_i|$ is odd. Since the sub-matrix inside the determinant of Eq.~(\ref{eq:otocexactline6}) must be square, we must have

\begin{align}
|\vec{B}^{\p}| + |\mathcal{V}^{\times g} \left(\vec{B} \right)| + |\va| &= |\vec{B}| + |\mathcal{V}^{\times g} \left(\vec{B}^{\p} \right)| + |\va|\\
|\mathcal{V}^{\times g} \left(\vec{B} \right)| - |\vec{B}| &= |\mathcal{V}^{\times g} \left(\vec{B}^{\p} \right)| - |\vec{B}^{\p}| \\
\frac{1}{2} \sum_{i = 1}^g \left[ |\mathcal{V}^{\times g} \left(\vb_i \right)| - |\vb_i| \right] &= \frac{1}{2} \sum_{i = 1}^{g} \left[|\mathcal{V}^{\times g} \left(\vb_i^{\p} \right)| - |\vb_i^{\p}|\right] \mathrm{,}
\end{align}

\noindent and thus

\begin{align}
(-1)^{\frac{1}{2} \sum_{i = 1}^g \left[ |\mathcal{V}^{\times g} \left(\vb_i \right)| - |\vb_i| \right]} &= i^{\frac{1}{2} \sum_{i = 1}^g \left\{\left[ |\mathcal{V}^{\times g} \left(\vb_i \right)| - |\vb_i| \right] + \left[ |\mathcal{V}^{\times g} \left(\vb^{\p}_i \right)| - |\vb^{\p}_i| \right]\right\}} \\
(-1)^{\frac{1}{2} \sum_{i = 1}^g \left[ |\mathcal{V}^{\times g} \left(\vb_i \right)| - |\vb_i| \right]} &= (-1)^{\sum_{i = 1}^g \left(\delta_{|\vb_i|, 3} + \delta_{|\vb^{\p}_i|, 3} \right)}i^{\frac{1}{2} \sum_{i = 1}^g \left[| |\mathcal{V}^{\times g} \left(\vb_i \right)| - |\vb_i|| + ||\mathcal{V}^{\times g} \left(\vb^{\p}_i \right)| - |\vb^{\p}_i| | \right]} \rm{,}
\end{align}

\noindent and the exponent on the factor of $i$ is the number of $\vb_i$ for which $|\vb_i|$ is odd, which cancels the corresponding factor of $-i$ from Eq.~(\ref{eq:interphase}).

\section*{Appendix D: Exact formula for Lightcone Boundary} \label{sec:exactboundary}

We want to calculate

\begin{align}
b_s^2(t) \equiv 
\begin{cases}
\sum_{\substack{\vb \ \mathrm{with} \ X_s I^{\otimes (n - s)} \ \\ \mathrm{or} \ Y_s I^{\otimes (n - s)} \ \mathrm{present}}} \det \left(\mathbf{u}_{\va \vb} \right)^2 & (s \geq \lfloor n/2 \rfloor) \\
\sum_{\substack{\vb \ \mathrm{with} \ I^{\otimes (s - 1)} X_s \ \\ \mathrm{or} \ I^{\otimes (s - 1)} Y_s \ \mathrm{present}}} \det \left(\mathbf{u}_{\va \vb} \right)^2 & (s \leq \lfloor n/2 \rfloor)\end{cases} \mathrm{.}
\label{eq:boundarydef}
\end{align}

\noindent As stated in the main text, we can apply the Jordan-Wigner transformation on the strings satisfying the condition in each sum to obtain

\begin{align}
b_s^2(t) = 
\begin{cases}
\sum_{\vb^{\p}} \left\{\det \left[\mathbf{u}_{\va \left(\vb^{\p}, 2s - 1\right)} \right]^2 + \det \left[\mathbf{u}_{\va \left(\vb^{\p}, 2s\right)} \right]^2\right\} & (s \geq \lfloor n/2 \rfloor) \\
\sum_{\vb^{\p}} \left\{\det \left[\mathbf{u}_{\va \left([2s - 2], 2s - 1, \vb^{\p}\right)} \right]^2 + \det \left[\mathbf{u}_{\va \left([2s - 2], 2s, \vb^{\p}\right)} \right]^2\right\} & (s \leq \lfloor n/2 \rfloor)\end{cases} \mathrm{.}
\label{eq:boundarydef}
\end{align}

\noindent Each of these sums is of the form in Eq.~(\ref{eq:modCBform}), which we can evaluate to obtain

\begin{equation}
-b_{s}^2(t) = \begin{cases} \det \begin{pmatrix} 0 & \mathbf{u}^{\mathrm{T}}_{2s - 1, \va} \\
\mathbf{u}_{\va, 2s - 1} & \mathbf{u}_{\va [2(s - 1)]} \mathbf{u}^{\mathrm{T}}_{[2(s - 1)] \va}
\end{pmatrix} + \det \begin{pmatrix} 0 & \mathbf{u}^{\mathrm{T}}_{2s, \va} \\
\mathbf{u}_{\va, 2s} & \mathbf{u}_{\va [2(s - 1)]} \mathbf{u}^{\mathrm{T}}_{[2(s - 1)] \va}
\end{pmatrix} & (s \geq \lfloor n/2 \rfloor) \\
 \det \begin{pmatrix} \mathbf{0}_{(2s - 1) \times (2s - 1)} & \mathbf{u}^{\mathrm{T}}_{([2(s - 1)], 2s - 1), \va} \\
\mathbf{u}_{\va, ([2(s - 1)], 2s - 1)} & \mathbf{u}_{\va \overline{[2s]}} \mathbf{u}^{\mathrm{T}}_{\overline{[2s]} \va}
\end{pmatrix} + \det \begin{pmatrix} \mathbf{0}_{(2s - 1) \times (2s - 1)} & \mathbf{u}^{\mathrm{T}}_{([2(s - 1)], 2s), \va} \\
\mathbf{u}_{\va, ([2(s - 1)], 2s)} & \mathbf{u}_{\va \overline{[2s]}} \mathbf{u}^{\mathrm{T}}_{\overline{[2s]} \va}
\end{pmatrix}& (s \leq \lfloor n/2 \rfloor)
\end{cases} \mathrm{.}
\label{eq:boundarycalc}
\end{equation}

\noindent We see that $b_s^2(t)$ can therefore be evaluated efficiently as the sum of only two determinants of polynomially-sized matrices.

\twocolumngrid


\begin{thebibliography}{55}%
\makeatletter
\providecommand \@ifxundefined [1]{%
 \@ifx{#1\undefined}
}%
\providecommand \@ifnum [1]{%
 \ifnum #1\expandafter \@firstoftwo
 \else \expandafter \@secondoftwo
 \fi
}%
\providecommand \@ifx [1]{%
 \ifx #1\expandafter \@firstoftwo
 \else \expandafter \@secondoftwo
 \fi
}%
\providecommand \natexlab [1]{#1}%
\providecommand \enquote  [1]{``#1''}%
\providecommand \bibnamefont  [1]{#1}%
\providecommand \bibfnamefont [1]{#1}%
\providecommand \citenamefont [1]{#1}%
\providecommand \href@noop [0]{\@secondoftwo}%
\providecommand \href [0]{\begingroup \@sanitize@url \@href}%
\providecommand \@href[1]{\@@startlink{#1}\@@href}%
\providecommand \@@href[1]{\endgroup#1\@@endlink}%
\providecommand \@sanitize@url [0]{\catcode `\\12\catcode `\$12\catcode
  `\&12\catcode `\#12\catcode `\^12\catcode `\_12\catcode `\%12\relax}%
\providecommand \@@startlink[1]{}%
\providecommand \@@endlink[0]{}%
\providecommand \url  [0]{\begingroup\@sanitize@url \@url }%
\providecommand \@url [1]{\endgroup\@href {#1}{\urlprefix }}%
\providecommand \urlprefix  [0]{URL }%
\providecommand \Eprint [0]{\href }%
\providecommand \doibase [0]{http://dx.doi.org/}%
\providecommand \selectlanguage [0]{\@gobble}%
\providecommand \bibinfo  [0]{\@secondoftwo}%
\providecommand \bibfield  [0]{\@secondoftwo}%
\providecommand \translation [1]{[#1]}%
\providecommand \BibitemOpen [0]{}%
\providecommand \bibitemStop [0]{}%
\providecommand \bibitemNoStop [0]{.\EOS\space}%
\providecommand \EOS [0]{\spacefactor3000\relax}%
\providecommand \BibitemShut  [1]{\csname bibitem#1\endcsname}%
\let\auto@bib@innerbib\@empty
\bibitem [{\citenamefont {Lieb}\ and\ \citenamefont
  {Robinson}(1972)}]{lieb1972finite}%
  \BibitemOpen
  \bibfield  {author} {\bibinfo {author} {\bibfnamefont {E.~H.}\ \bibnamefont
  {Lieb}}\ and\ \bibinfo {author} {\bibfnamefont {D.~W.}\ \bibnamefont
  {Robinson}},\ }\href {\doibase 10.1007/BF01645779} {\bibfield  {journal}
  {\bibinfo  {journal} {Communications in Mathematical Physics}\ }\textbf
  {\bibinfo {volume} {28}},\ \bibinfo {pages} {251} (\bibinfo {year}
  {1972})}\BibitemShut {NoStop}%
\bibitem [{\citenamefont {Gornyi}\ \emph {et~al.}(2005)\citenamefont {Gornyi},
  \citenamefont {Mirlin},\ and\ \citenamefont
  {Polyakov}}]{gornyi2005interacting}%
  \BibitemOpen
  \bibfield  {author} {\bibinfo {author} {\bibfnamefont {I.~V.}\ \bibnamefont
  {Gornyi}}, \bibinfo {author} {\bibfnamefont {A.~D.}\ \bibnamefont {Mirlin}},
  \ and\ \bibinfo {author} {\bibfnamefont {D.~G.}\ \bibnamefont {Polyakov}},\
  }\href {\doibase 10.1103/PhysRevLett.95.206603} {\bibfield  {journal}
  {\bibinfo  {journal} {Phys. Rev. Lett.}\ }\textbf {\bibinfo {volume} {95}},\
  \bibinfo {pages} {206603} (\bibinfo {year} {2005})}\BibitemShut {NoStop}%
\bibitem [{\citenamefont {Basko}\ \emph {et~al.}(2006)\citenamefont {Basko},
  \citenamefont {Aleiner},\ and\ \citenamefont {Altshuler}}]{basko2006metal}%
  \BibitemOpen
  \bibfield  {author} {\bibinfo {author} {\bibfnamefont {D.}~\bibnamefont
  {Basko}}, \bibinfo {author} {\bibfnamefont {I.}~\bibnamefont {Aleiner}}, \
  and\ \bibinfo {author} {\bibfnamefont {B.}~\bibnamefont {Altshuler}},\ }\href
  {\doibase https://doi.org/10.1016/j.aop.2005.11.014} {\bibfield  {journal}
  {\bibinfo  {journal} {Annals of Physics}\ }\textbf {\bibinfo {volume}
  {321}},\ \bibinfo {pages} {1126 } (\bibinfo {year} {2006})}\BibitemShut
  {NoStop}%
\bibitem [{\citenamefont {Imbrie}(2016)}]{imbrie2016mbl}%
  \BibitemOpen
  \bibfield  {author} {\bibinfo {author} {\bibfnamefont {J.~Z.}\ \bibnamefont
  {Imbrie}},\ }\href {\doibase 10.1007/s10955-016-1508-x} {\bibfield  {journal}
  {\bibinfo  {journal} {Journal of Statistical Physics}\ }\textbf {\bibinfo
  {volume} {163}},\ \bibinfo {pages} {998} (\bibinfo {year}
  {2016})}\BibitemShut {NoStop}%
\bibitem [{\citenamefont {Serbyn}\ \emph {et~al.}(2013)\citenamefont {Serbyn},
  \citenamefont {Papi\ifmmode~\acute{c}\else \'{c}\fi{}},\ and\ \citenamefont
  {Abanin}}]{serbyn2013local}%
  \BibitemOpen
  \bibfield  {author} {\bibinfo {author} {\bibfnamefont {M.}~\bibnamefont
  {Serbyn}}, \bibinfo {author} {\bibfnamefont {Z.}~\bibnamefont
  {Papi\ifmmode~\acute{c}\else \'{c}\fi{}}}, \ and\ \bibinfo {author}
  {\bibfnamefont {D.~A.}\ \bibnamefont {Abanin}},\ }\href {\doibase
  10.1103/PhysRevLett.111.127201} {\bibfield  {journal} {\bibinfo  {journal}
  {Phys. Rev. Lett.}\ }\textbf {\bibinfo {volume} {111}},\ \bibinfo {pages}
  {127201} (\bibinfo {year} {2013})}\BibitemShut {NoStop}%
\bibitem [{\citenamefont {Huse}\ \emph {et~al.}(2014)\citenamefont {Huse},
  \citenamefont {Nandkishore},\ and\ \citenamefont
  {Oganesyan}}]{huse2014phenomenology}%
  \BibitemOpen
  \bibfield  {author} {\bibinfo {author} {\bibfnamefont {D.~A.}\ \bibnamefont
  {Huse}}, \bibinfo {author} {\bibfnamefont {R.}~\bibnamefont {Nandkishore}}, \
  and\ \bibinfo {author} {\bibfnamefont {V.}~\bibnamefont {Oganesyan}},\ }\href
  {\doibase 10.1103/PhysRevB.90.174202} {\bibfield  {journal} {\bibinfo
  {journal} {Phys. Rev. B}\ }\textbf {\bibinfo {volume} {90}},\ \bibinfo
  {pages} {174202} (\bibinfo {year} {2014})}\BibitemShut {NoStop}%
\bibitem [{\citenamefont {Swingle}\ and\ \citenamefont
  {Chowdhury}(2017)}]{chowdhury2016slow}%
  \BibitemOpen
  \bibfield  {author} {\bibinfo {author} {\bibfnamefont {B.}~\bibnamefont
  {Swingle}}\ and\ \bibinfo {author} {\bibfnamefont {D.}~\bibnamefont
  {Chowdhury}},\ }\href {\doibase 10.1103/PhysRevB.95.060201} {\bibfield
  {journal} {\bibinfo  {journal} {Phys. Rev. B}\ }\textbf {\bibinfo {volume}
  {95}},\ \bibinfo {pages} {060201} (\bibinfo {year} {2017})}\BibitemShut
  {NoStop}%
\bibitem [{\citenamefont {Deng}\ \emph {et~al.}(2017)\citenamefont {Deng},
  \citenamefont {Li}, \citenamefont {Pixley}, \citenamefont {Wu},\ and\
  \citenamefont {Das~Sarma}}]{deng2017logarithmic}%
  \BibitemOpen
  \bibfield  {author} {\bibinfo {author} {\bibfnamefont {D.-L.}\ \bibnamefont
  {Deng}}, \bibinfo {author} {\bibfnamefont {X.}~\bibnamefont {Li}}, \bibinfo
  {author} {\bibfnamefont {J.~H.}\ \bibnamefont {Pixley}}, \bibinfo {author}
  {\bibfnamefont {Y.-L.}\ \bibnamefont {Wu}}, \ and\ \bibinfo {author}
  {\bibfnamefont {S.}~\bibnamefont {Das~Sarma}},\ }\href {\doibase
  10.1103/PhysRevB.95.024202} {\bibfield  {journal} {\bibinfo  {journal} {Phys.
  Rev. B}\ }\textbf {\bibinfo {volume} {95}},\ \bibinfo {pages} {024202}
  (\bibinfo {year} {2017})}\BibitemShut {NoStop}%
\bibitem [{\citenamefont {Nanduri}\ \emph {et~al.}(2014)\citenamefont
  {Nanduri}, \citenamefont {Kim},\ and\ \citenamefont
  {Huse}}]{nanduri2014entanglement}%
  \BibitemOpen
  \bibfield  {author} {\bibinfo {author} {\bibfnamefont {A.}~\bibnamefont
  {Nanduri}}, \bibinfo {author} {\bibfnamefont {H.}~\bibnamefont {Kim}}, \ and\
  \bibinfo {author} {\bibfnamefont {D.~A.}\ \bibnamefont {Huse}},\ }\href
  {\doibase 10.1103/PhysRevB.90.064201} {\bibfield  {journal} {\bibinfo
  {journal} {Phys. Rev. B}\ }\textbf {\bibinfo {volume} {90}},\ \bibinfo
  {pages} {064201} (\bibinfo {year} {2014})}\BibitemShut {NoStop}%
\bibitem [{\citenamefont {Bravyi}\ \emph {et~al.}(2006)\citenamefont {Bravyi},
  \citenamefont {Hastings},\ and\ \citenamefont
  {Verstraete}}]{bravyi2006liebrobinson}%
  \BibitemOpen
  \bibfield  {author} {\bibinfo {author} {\bibfnamefont {S.}~\bibnamefont
  {Bravyi}}, \bibinfo {author} {\bibfnamefont {M.~B.}\ \bibnamefont
  {Hastings}}, \ and\ \bibinfo {author} {\bibfnamefont {F.}~\bibnamefont
  {Verstraete}},\ }\href {\doibase 10.1103/PhysRevLett.97.050401} {\bibfield
  {journal} {\bibinfo  {journal} {Phys. Rev. Lett.}\ }\textbf {\bibinfo
  {volume} {97}},\ \bibinfo {pages} {050401} (\bibinfo {year}
  {2006})}\BibitemShut {NoStop}%
\bibitem [{\citenamefont {\ifmmode \check{Z}\else
  \v{Z}\fi{}nidari\ifmmode~\check{c}\else \v{c}\fi{}}\ \emph
  {et~al.}(2008)\citenamefont {\ifmmode \check{Z}\else
  \v{Z}\fi{}nidari\ifmmode~\check{c}\else \v{c}\fi{}}, \citenamefont {Prosen},\
  and\ \citenamefont {Prelov\ifmmode~\check{s}\else
  \v{s}\fi{}ek}}]{znidaric2008manybody}%
  \BibitemOpen
  \bibfield  {author} {\bibinfo {author} {\bibfnamefont {M.}~\bibnamefont
  {\ifmmode \check{Z}\else \v{Z}\fi{}nidari\ifmmode~\check{c}\else
  \v{c}\fi{}}}, \bibinfo {author} {\bibfnamefont {T.~c.~v.}\ \bibnamefont
  {Prosen}}, \ and\ \bibinfo {author} {\bibfnamefont {P.}~\bibnamefont
  {Prelov\ifmmode~\check{s}\else \v{s}\fi{}ek}},\ }\href {\doibase
  10.1103/PhysRevB.77.064426} {\bibfield  {journal} {\bibinfo  {journal} {Phys.
  Rev. B}\ }\textbf {\bibinfo {volume} {77}},\ \bibinfo {pages} {064426}
  (\bibinfo {year} {2008})}\BibitemShut {NoStop}%
\bibitem [{\citenamefont {Bardarson}\ \emph {et~al.}(2012)\citenamefont
  {Bardarson}, \citenamefont {Pollmann},\ and\ \citenamefont
  {Moore}}]{bardarson2012unbounded}%
  \BibitemOpen
  \bibfield  {author} {\bibinfo {author} {\bibfnamefont {J.~H.}\ \bibnamefont
  {Bardarson}}, \bibinfo {author} {\bibfnamefont {F.}~\bibnamefont {Pollmann}},
  \ and\ \bibinfo {author} {\bibfnamefont {J.~E.}\ \bibnamefont {Moore}},\
  }\href {\doibase 10.1103/PhysRevLett.109.017202} {\bibfield  {journal}
  {\bibinfo  {journal} {Phys. Rev. Lett.}\ }\textbf {\bibinfo {volume} {109}},\
  \bibinfo {pages} {017202} (\bibinfo {year} {2012})}\BibitemShut {NoStop}%
\bibitem [{\citenamefont {{Kim}}\ \emph {et~al.}(2014)\citenamefont {{Kim}},
  \citenamefont {{Chandran}},\ and\ \citenamefont {{Abanin}}}]{kim2014local}%
  \BibitemOpen
  \bibfield  {author} {\bibinfo {author} {\bibfnamefont {I.~H.}\ \bibnamefont
  {{Kim}}}, \bibinfo {author} {\bibfnamefont {A.}~\bibnamefont {{Chandran}}}, \
  and\ \bibinfo {author} {\bibfnamefont {D.~A.}\ \bibnamefont {{Abanin}}},\
  }\href@noop {} {\bibfield  {journal} {\bibinfo  {journal} {ArXiv e-prints}\ }
  (\bibinfo {year} {2014})},\ \Eprint {http://arxiv.org/abs/1412.3073}
  {arXiv:1412.3073 [cond-mat.dis-nn]} \BibitemShut {NoStop}%
\bibitem [{\citenamefont {Ba\~nuls}\ \emph {et~al.}(2017)\citenamefont
  {Ba\~nuls}, \citenamefont {Yao}, \citenamefont {Choi}, \citenamefont
  {Lukin},\ and\ \citenamefont {Cirac}}]{cirac2017dynamics}%
  \BibitemOpen
  \bibfield  {author} {\bibinfo {author} {\bibfnamefont {M.~C.}\ \bibnamefont
  {Ba\~nuls}}, \bibinfo {author} {\bibfnamefont {N.~Y.}\ \bibnamefont {Yao}},
  \bibinfo {author} {\bibfnamefont {S.}~\bibnamefont {Choi}}, \bibinfo {author}
  {\bibfnamefont {M.~D.}\ \bibnamefont {Lukin}}, \ and\ \bibinfo {author}
  {\bibfnamefont {J.~I.}\ \bibnamefont {Cirac}},\ }\href {\doibase
  10.1103/PhysRevB.96.174201} {\bibfield  {journal} {\bibinfo  {journal} {Phys.
  Rev. B}\ }\textbf {\bibinfo {volume} {96}},\ \bibinfo {pages} {174201}
  (\bibinfo {year} {2017})}\BibitemShut {NoStop}%
\bibitem [{\citenamefont {{Chandran}}\ and\ \citenamefont
  {{Laumann}}(2015)}]{chandran2015semiclassical}%
  \BibitemOpen
  \bibfield  {author} {\bibinfo {author} {\bibfnamefont {A.}~\bibnamefont
  {{Chandran}}}\ and\ \bibinfo {author} {\bibfnamefont {C.~R.}\ \bibnamefont
  {{Laumann}}},\ }\href {\doibase 10.1103/PhysRevB.92.024301} {\bibfield
  {journal} {\bibinfo  {journal} {\prb}\ }\textbf {\bibinfo {volume} {92}},\
  \bibinfo {eid} {024301} (\bibinfo {year} {2015})},\ \Eprint
  {http://arxiv.org/abs/1501.01971} {arXiv:1501.01971 [cond-mat.dis-nn]}
  \BibitemShut {NoStop}%
\bibitem [{\citenamefont {Keser}\ \emph {et~al.}(2016)\citenamefont {Keser},
  \citenamefont {Ganeshan}, \citenamefont {Refael},\ and\ \citenamefont
  {Galitski}}]{keser2016dynamical}%
  \BibitemOpen
  \bibfield  {author} {\bibinfo {author} {\bibfnamefont {A.~C.}\ \bibnamefont
  {Keser}}, \bibinfo {author} {\bibfnamefont {S.}~\bibnamefont {Ganeshan}},
  \bibinfo {author} {\bibfnamefont {G.}~\bibnamefont {Refael}}, \ and\ \bibinfo
  {author} {\bibfnamefont {V.}~\bibnamefont {Galitski}},\ }\href {\doibase
  10.1103/PhysRevB.94.085120} {\bibfield  {journal} {\bibinfo  {journal} {Phys.
  Rev. B}\ }\textbf {\bibinfo {volume} {94}},\ \bibinfo {pages} {085120}
  (\bibinfo {year} {2016})}\BibitemShut {NoStop}%
\bibitem [{\citenamefont {Abanin}\ \emph {et~al.}(2016)\citenamefont {Abanin},
  \citenamefont {Roeck},\ and\ \citenamefont {Huveneers}}]{abanin2016theory}%
  \BibitemOpen
  \bibfield  {author} {\bibinfo {author} {\bibfnamefont {D.~A.}\ \bibnamefont
  {Abanin}}, \bibinfo {author} {\bibfnamefont {W.~D.}\ \bibnamefont {Roeck}}, \
  and\ \bibinfo {author} {\bibfnamefont {F.}~\bibnamefont {Huveneers}},\ }\href
  {\doibase https://doi.org/10.1016/j.aop.2016.03.010} {\bibfield  {journal}
  {\bibinfo  {journal} {Annals of Physics}\ }\textbf {\bibinfo {volume}
  {372}},\ \bibinfo {pages} {1 } (\bibinfo {year} {2016})}\BibitemShut
  {NoStop}%
\bibitem [{\citenamefont {Ponte}\ \emph
  {et~al.}(2015{\natexlab{a}})\citenamefont {Ponte}, \citenamefont
  {Papi\ifmmode~\acute{c}\else \'{c}\fi{}}, \citenamefont {Huveneers},\ and\
  \citenamefont {Abanin}}]{ponte2015mbl}%
  \BibitemOpen
  \bibfield  {author} {\bibinfo {author} {\bibfnamefont {P.}~\bibnamefont
  {Ponte}}, \bibinfo {author} {\bibfnamefont {Z.}~\bibnamefont
  {Papi\ifmmode~\acute{c}\else \'{c}\fi{}}}, \bibinfo {author} {\bibfnamefont
  {F.~m.~c.}\ \bibnamefont {Huveneers}}, \ and\ \bibinfo {author}
  {\bibfnamefont {D.~A.}\ \bibnamefont {Abanin}},\ }\href {\doibase
  10.1103/PhysRevLett.114.140401} {\bibfield  {journal} {\bibinfo  {journal}
  {Phys. Rev. Lett.}\ }\textbf {\bibinfo {volume} {114}},\ \bibinfo {pages}
  {140401} (\bibinfo {year} {2015}{\natexlab{a}})}\BibitemShut {NoStop}%
\bibitem [{\citenamefont {Ponte}\ \emph
  {et~al.}(2015{\natexlab{b}})\citenamefont {Ponte}, \citenamefont {Chandran},
  \citenamefont {Papić},\ and\ \citenamefont
  {Abanin}}]{ponte2015periodically}%
  \BibitemOpen
  \bibfield  {author} {\bibinfo {author} {\bibfnamefont {P.}~\bibnamefont
  {Ponte}}, \bibinfo {author} {\bibfnamefont {A.}~\bibnamefont {Chandran}},
  \bibinfo {author} {\bibfnamefont {Z.}~\bibnamefont {Papić}}, \ and\ \bibinfo
  {author} {\bibfnamefont {D.~A.}\ \bibnamefont {Abanin}},\ }\href {\doibase
  https://doi.org/10.1016/j.aop.2014.11.008} {\bibfield  {journal} {\bibinfo
  {journal} {Annals of Physics}\ }\textbf {\bibinfo {volume} {353}},\ \bibinfo
  {pages} {196 } (\bibinfo {year} {2015}{\natexlab{b}})}\BibitemShut {NoStop}%
\bibitem [{\citenamefont {Mierzejewski}\ \emph {et~al.}(2017)\citenamefont
  {Mierzejewski}, \citenamefont {Giergiel},\ and\ \citenamefont
  {Sacha}}]{mierzejewski2017mbl}%
  \BibitemOpen
  \bibfield  {author} {\bibinfo {author} {\bibfnamefont {M.}~\bibnamefont
  {Mierzejewski}}, \bibinfo {author} {\bibfnamefont {K.}~\bibnamefont
  {Giergiel}}, \ and\ \bibinfo {author} {\bibfnamefont {K.}~\bibnamefont
  {Sacha}},\ }\href {\doibase 10.1103/PhysRevB.96.140201} {\bibfield  {journal}
  {\bibinfo  {journal} {Phys. Rev. B}\ }\textbf {\bibinfo {volume} {96}},\
  \bibinfo {pages} {140201} (\bibinfo {year} {2017})}\BibitemShut {NoStop}%
\bibitem [{\citenamefont {{Kitaev}}(2015)}]{kitaev2015kitp}%
  \BibitemOpen
  \bibfield  {author} {\bibinfo {author} {\bibfnamefont {A.}~\bibnamefont
  {{Kitaev}}},\ }\href@noop {} {\enquote {\bibinfo {title} {A simple model of
  quantum holography},}\ } (\bibinfo {year} {2015}),\ \bibinfo {note} {{KITP}
  strings seminar and Entanglement 2015 program}\BibitemShut {NoStop}%
\bibitem [{\citenamefont {{Gullans}}\ and\ \citenamefont
  {{Huse}}(2018)}]{gullans2018entanglement}%
  \BibitemOpen
  \bibfield  {author} {\bibinfo {author} {\bibfnamefont {M.~J.}\ \bibnamefont
  {{Gullans}}}\ and\ \bibinfo {author} {\bibfnamefont {D.~A.}\ \bibnamefont
  {{Huse}}},\ }\href@noop {} {\bibfield  {journal} {\bibinfo  {journal} {ArXiv
  e-prints}\ } (\bibinfo {year} {2018})},\ \Eprint
  {http://arxiv.org/abs/1804.00010} {arXiv:1804.00010 [cond-mat.stat-mech]}
  \BibitemShut {NoStop}%
\bibitem [{\citenamefont {{Zhou}}\ and\ \citenamefont
  {{Nahum}}(2018)}]{zhou2018emergent}%
  \BibitemOpen
  \bibfield  {author} {\bibinfo {author} {\bibfnamefont {T.}~\bibnamefont
  {{Zhou}}}\ and\ \bibinfo {author} {\bibfnamefont {A.}~\bibnamefont
  {{Nahum}}},\ }\href@noop {} {\bibfield  {journal} {\bibinfo  {journal} {ArXiv
  e-prints}\ } (\bibinfo {year} {2018})},\ \Eprint
  {http://arxiv.org/abs/1804.09737} {arXiv:1804.09737 [cond-mat.stat-mech]}
  \BibitemShut {NoStop}%
\bibitem [{\citenamefont {{Zhou}}\ and\ \citenamefont
  {{Chen}}(2018)}]{zhou2018operator}%
  \BibitemOpen
  \bibfield  {author} {\bibinfo {author} {\bibfnamefont {T.}~\bibnamefont
  {{Zhou}}}\ and\ \bibinfo {author} {\bibfnamefont {X.}~\bibnamefont
  {{Chen}}},\ }\href@noop {} {\bibfield  {journal} {\bibinfo  {journal} {ArXiv
  e-prints}\ } (\bibinfo {year} {2018})},\ \Eprint
  {http://arxiv.org/abs/1805.09307} {arXiv:1805.09307 [cond-mat.str-el]}
  \BibitemShut {NoStop}%
\bibitem [{\citenamefont {{Jonay}}\ \emph {et~al.}(2018)\citenamefont
  {{Jonay}}, \citenamefont {{Huse}},\ and\ \citenamefont
  {{Nahum}}}]{jonay2018coarsegrained}%
  \BibitemOpen
  \bibfield  {author} {\bibinfo {author} {\bibfnamefont {C.}~\bibnamefont
  {{Jonay}}}, \bibinfo {author} {\bibfnamefont {D.~A.}\ \bibnamefont {{Huse}}},
  \ and\ \bibinfo {author} {\bibfnamefont {A.}~\bibnamefont {{Nahum}}},\
  }\href@noop {} {\bibfield  {journal} {\bibinfo  {journal} {ArXiv e-prints}\ }
  (\bibinfo {year} {2018})},\ \Eprint {http://arxiv.org/abs/1803.00089}
  {arXiv:1803.00089 [cond-mat.stat-mech]} \BibitemShut {NoStop}%
\bibitem [{\citenamefont {{Xu}}\ and\ \citenamefont
  {{Swingle}}(2018{\natexlab{a}})}]{xu2018locality}%
  \BibitemOpen
  \bibfield  {author} {\bibinfo {author} {\bibfnamefont {S.}~\bibnamefont
  {{Xu}}}\ and\ \bibinfo {author} {\bibfnamefont {B.}~\bibnamefont
  {{Swingle}}},\ }\href@noop {} {\bibfield  {journal} {\bibinfo  {journal}
  {ArXiv e-prints}\ } (\bibinfo {year} {2018}{\natexlab{a}})},\ \Eprint
  {http://arxiv.org/abs/1805.05376} {arXiv:1805.05376 [cond-mat.str-el]}
  \BibitemShut {NoStop}%
\bibitem [{\citenamefont {{S{\"u}nderhauf}}\ \emph {et~al.}(2018)\citenamefont
  {{S{\"u}nderhauf}}, \citenamefont {{P{\'e}rez-Garc{\'{\i}}a}}, \citenamefont
  {{Huse}}, \citenamefont {{Schuch}},\ and\ \citenamefont
  {{Cirac}}}]{sunderhauf2018localisation}%
  \BibitemOpen
  \bibfield  {author} {\bibinfo {author} {\bibfnamefont {C.}~\bibnamefont
  {{S{\"u}nderhauf}}}, \bibinfo {author} {\bibfnamefont {D.}~\bibnamefont
  {{P{\'e}rez-Garc{\'{\i}}a}}}, \bibinfo {author} {\bibfnamefont {D.~A.}\
  \bibnamefont {{Huse}}}, \bibinfo {author} {\bibfnamefont {N.}~\bibnamefont
  {{Schuch}}}, \ and\ \bibinfo {author} {\bibfnamefont {J.~I.}\ \bibnamefont
  {{Cirac}}},\ }\href@noop {} {\bibfield  {journal} {\bibinfo  {journal} {ArXiv
  e-prints}\ } (\bibinfo {year} {2018})},\ \Eprint
  {http://arxiv.org/abs/1805.08487} {arXiv:1805.08487 [cond-mat.stat-mech]}
  \BibitemShut {NoStop}%
\bibitem [{\citenamefont {Nahum}\ \emph {et~al.}(2018)\citenamefont {Nahum},
  \citenamefont {Vijay},\ and\ \citenamefont {Haah}}]{nahum2018operator}%
  \BibitemOpen
  \bibfield  {author} {\bibinfo {author} {\bibfnamefont {A.}~\bibnamefont
  {Nahum}}, \bibinfo {author} {\bibfnamefont {S.}~\bibnamefont {Vijay}}, \ and\
  \bibinfo {author} {\bibfnamefont {J.}~\bibnamefont {Haah}},\ }\href {\doibase
  10.1103/PhysRevX.8.021014} {\bibfield  {journal} {\bibinfo  {journal} {Phys.
  Rev. X}\ }\textbf {\bibinfo {volume} {8}},\ \bibinfo {pages} {021014}
  (\bibinfo {year} {2018})}\BibitemShut {NoStop}%
\bibitem [{\citenamefont {von Keyserlingk}\ \emph {et~al.}(2018)\citenamefont
  {von Keyserlingk}, \citenamefont {Rakovszky}, \citenamefont {Pollmann},\ and\
  \citenamefont {Sondhi}}]{keyserlingk2017operator}%
  \BibitemOpen
  \bibfield  {author} {\bibinfo {author} {\bibfnamefont {C.~W.}\ \bibnamefont
  {von Keyserlingk}}, \bibinfo {author} {\bibfnamefont {T.}~\bibnamefont
  {Rakovszky}}, \bibinfo {author} {\bibfnamefont {F.}~\bibnamefont {Pollmann}},
  \ and\ \bibinfo {author} {\bibfnamefont {S.~L.}\ \bibnamefont {Sondhi}},\
  }\href {\doibase 10.1103/PhysRevX.8.021013} {\bibfield  {journal} {\bibinfo
  {journal} {Phys. Rev. X}\ }\textbf {\bibinfo {volume} {8}},\ \bibinfo {pages}
  {021013} (\bibinfo {year} {2018})}\BibitemShut {NoStop}%
\bibitem [{\citenamefont {{Rakovszky}}\ \emph {et~al.}(2017)\citenamefont
  {{Rakovszky}}, \citenamefont {{Pollmann}},\ and\ \citenamefont {{von
  Keyserlingk}}}]{rakovszky2018diffusive}%
  \BibitemOpen
  \bibfield  {author} {\bibinfo {author} {\bibfnamefont {T.}~\bibnamefont
  {{Rakovszky}}}, \bibinfo {author} {\bibfnamefont {F.}~\bibnamefont
  {{Pollmann}}}, \ and\ \bibinfo {author} {\bibfnamefont {C.~W.}\ \bibnamefont
  {{von Keyserlingk}}},\ }\href@noop {} {\bibfield  {journal} {\bibinfo
  {journal} {ArXiv e-prints}\ } (\bibinfo {year} {2017})},\ \Eprint
  {http://arxiv.org/abs/1710.09827} {arXiv:1710.09827 [cond-mat.stat-mech]}
  \BibitemShut {NoStop}%
\bibitem [{\citenamefont {{Khemani}}\ \emph {et~al.}(2017)\citenamefont
  {{Khemani}}, \citenamefont {{Vishwanath}},\ and\ \citenamefont
  {{Huse}}}]{khemani2017operator}%
  \BibitemOpen
  \bibfield  {author} {\bibinfo {author} {\bibfnamefont {V.}~\bibnamefont
  {{Khemani}}}, \bibinfo {author} {\bibfnamefont {A.}~\bibnamefont
  {{Vishwanath}}}, \ and\ \bibinfo {author} {\bibfnamefont {D.~A.}\
  \bibnamefont {{Huse}}},\ }\href@noop {} {\bibfield  {journal} {\bibinfo
  {journal} {ArXiv e-prints}\ } (\bibinfo {year} {2017})},\ \Eprint
  {http://arxiv.org/abs/1710.09835} {arXiv:1710.09835 [cond-mat.stat-mech]}
  \BibitemShut {NoStop}%
\bibitem [{\citenamefont {{Gharibyan}}\ \emph {et~al.}(2018)\citenamefont
  {{Gharibyan}}, \citenamefont {{Hanada}}, \citenamefont {{Shenker}},\ and\
  \citenamefont {{Tezuka}}}]{gharibyan2018onset}%
  \BibitemOpen
  \bibfield  {author} {\bibinfo {author} {\bibfnamefont {H.}~\bibnamefont
  {{Gharibyan}}}, \bibinfo {author} {\bibfnamefont {M.}~\bibnamefont
  {{Hanada}}}, \bibinfo {author} {\bibfnamefont {S.~H.}\ \bibnamefont
  {{Shenker}}}, \ and\ \bibinfo {author} {\bibfnamefont {M.}~\bibnamefont
  {{Tezuka}}},\ }\href@noop {} {\bibfield  {journal} {\bibinfo  {journal}
  {ArXiv e-prints}\ } (\bibinfo {year} {2018})},\ \Eprint
  {http://arxiv.org/abs/1803.08050} {arXiv:1803.08050 [hep-th]} \BibitemShut
  {NoStop}%
\bibitem [{\citenamefont {Kos}\ \emph {et~al.}(2018)\citenamefont {Kos},
  \citenamefont {Ljubotina},\ and\ \citenamefont {Prosen}}]{kos2018manybody}%
  \BibitemOpen
  \bibfield  {author} {\bibinfo {author} {\bibfnamefont {P.}~\bibnamefont
  {Kos}}, \bibinfo {author} {\bibfnamefont {M.}~\bibnamefont {Ljubotina}}, \
  and\ \bibinfo {author} {\bibfnamefont {T.~c.~v.}\ \bibnamefont {Prosen}},\
  }\href {\doibase 10.1103/PhysRevX.8.021062} {\bibfield  {journal} {\bibinfo
  {journal} {Phys. Rev. X}\ }\textbf {\bibinfo {volume} {8}},\ \bibinfo {pages}
  {021062} (\bibinfo {year} {2018})}\BibitemShut {NoStop}%
\bibitem [{\citenamefont {{Chen}}\ and\ \citenamefont
  {{Zhou}}(2018)}]{chen2018operator}%
  \BibitemOpen
  \bibfield  {author} {\bibinfo {author} {\bibfnamefont {X.}~\bibnamefont
  {{Chen}}}\ and\ \bibinfo {author} {\bibfnamefont {T.}~\bibnamefont
  {{Zhou}}},\ }\href@noop {} {\bibfield  {journal} {\bibinfo  {journal} {ArXiv
  e-prints}\ } (\bibinfo {year} {2018})},\ \Eprint
  {http://arxiv.org/abs/1804.08655} {arXiv:1804.08655 [cond-mat.str-el]}
  \BibitemShut {NoStop}%
\bibitem [{\citenamefont {Fan}\ \emph {et~al.}(2017)\citenamefont {Fan},
  \citenamefont {Zhang}, \citenamefont {Shen},\ and\ \citenamefont
  {Zhai}}]{fan2017oto}%
  \BibitemOpen
  \bibfield  {author} {\bibinfo {author} {\bibfnamefont {R.}~\bibnamefont
  {Fan}}, \bibinfo {author} {\bibfnamefont {P.}~\bibnamefont {Zhang}}, \bibinfo
  {author} {\bibfnamefont {H.}~\bibnamefont {Shen}}, \ and\ \bibinfo {author}
  {\bibfnamefont {H.}~\bibnamefont {Zhai}},\ }\href {\doibase
  https://www.sciencedirect.com/science/article/pii/S2095927317301925}
  {\bibfield  {journal} {\bibinfo  {journal} {Science Bulletin}\ }\textbf
  {\bibinfo {volume} {62}},\ \bibinfo {pages} {707 } (\bibinfo {year}
  {2017})}\BibitemShut {NoStop}%
\bibitem [{\citenamefont {He}\ and\ \citenamefont
  {Lu}(2017)}]{he2016characterizing}%
  \BibitemOpen
  \bibfield  {author} {\bibinfo {author} {\bibfnamefont {R.-Q.}\ \bibnamefont
  {He}}\ and\ \bibinfo {author} {\bibfnamefont {Z.-Y.}\ \bibnamefont {Lu}},\
  }\href {\doibase 10.1103/PhysRevB.95.054201} {\bibfield  {journal} {\bibinfo
  {journal} {Phys. Rev. B}\ }\textbf {\bibinfo {volume} {95}},\ \bibinfo
  {pages} {054201} (\bibinfo {year} {2017})}\BibitemShut {NoStop}%
\bibitem [{\citenamefont {{Chen}}(2016)}]{chen2016universal}%
  \BibitemOpen
  \bibfield  {author} {\bibinfo {author} {\bibfnamefont {Y.}~\bibnamefont
  {{Chen}}},\ }\href@noop {} {\bibfield  {journal} {\bibinfo  {journal} {ArXiv
  e-prints}\ } (\bibinfo {year} {2016})},\ \Eprint
  {http://arxiv.org/abs/1608.02765} {arXiv:1608.02765 [cond-mat.dis-nn]}
  \BibitemShut {NoStop}%
\bibitem [{\citenamefont {Xiao}\ \emph {et~al.}()\citenamefont {Xiao},
  \citenamefont {Tianci}, \citenamefont {A.},\ and\ \citenamefont
  {Eduardo}}]{xiao2016oto}%
  \BibitemOpen
  \bibfield  {author} {\bibinfo {author} {\bibfnamefont {C.}~\bibnamefont
  {Xiao}}, \bibinfo {author} {\bibfnamefont {Z.}~\bibnamefont {Tianci}},
  \bibinfo {author} {\bibfnamefont {H.~D.}\ \bibnamefont {A.}}, \ and\ \bibinfo
  {author} {\bibfnamefont {F.}~\bibnamefont {Eduardo}},\ }\href {\doibase
  10.1002/andp.201600332} {\bibfield  {journal} {\bibinfo  {journal} {Annalen
  der Physik}\ }\textbf {\bibinfo {volume} {529}},\ \bibinfo {pages}
  {1600332}}\BibitemShut {NoStop}%
\bibitem [{\citenamefont {Slagle}\ \emph {et~al.}(2017)\citenamefont {Slagle},
  \citenamefont {Bi}, \citenamefont {You},\ and\ \citenamefont
  {Xu}}]{slagle2017oto}%
  \BibitemOpen
  \bibfield  {author} {\bibinfo {author} {\bibfnamefont {K.}~\bibnamefont
  {Slagle}}, \bibinfo {author} {\bibfnamefont {Z.}~\bibnamefont {Bi}}, \bibinfo
  {author} {\bibfnamefont {Y.-Z.}\ \bibnamefont {You}}, \ and\ \bibinfo
  {author} {\bibfnamefont {C.}~\bibnamefont {Xu}},\ }\href {\doibase
  10.1103/PhysRevB.95.165136} {\bibfield  {journal} {\bibinfo  {journal} {Phys.
  Rev. B}\ }\textbf {\bibinfo {volume} {95}},\ \bibinfo {pages} {165136}
  (\bibinfo {year} {2017})}\BibitemShut {NoStop}%
\bibitem [{\citenamefont {Yichen}\ \emph {et~al.}()\citenamefont {Yichen},
  \citenamefont {Yong‐Liang},\ and\ \citenamefont {Xie}}]{huang2016oto}%
  \BibitemOpen
  \bibfield  {author} {\bibinfo {author} {\bibfnamefont {H.}~\bibnamefont
  {Yichen}}, \bibinfo {author} {\bibfnamefont {Z.}~\bibnamefont
  {Yong‐Liang}}, \ and\ \bibinfo {author} {\bibfnamefont {C.}~\bibnamefont
  {Xie}},\ }\href {\doibase 10.1002/andp.201600318} {\bibfield  {journal}
  {\bibinfo  {journal} {Annalen der Physik}\ }\textbf {\bibinfo {volume}
  {529}},\ \bibinfo {pages} {1600318}}\BibitemShut {NoStop}%
\bibitem [{\citenamefont {{Xu}}\ and\ \citenamefont
  {{Swingle}}(2018{\natexlab{b}})}]{xu2018accessing}%
  \BibitemOpen
  \bibfield  {author} {\bibinfo {author} {\bibfnamefont {S.}~\bibnamefont
  {{Xu}}}\ and\ \bibinfo {author} {\bibfnamefont {B.}~\bibnamefont
  {{Swingle}}},\ }\href@noop {} {\bibfield  {journal} {\bibinfo  {journal}
  {ArXiv e-prints}\ } (\bibinfo {year} {2018}{\natexlab{b}})},\ \Eprint
  {http://arxiv.org/abs/1802.00801} {arXiv:1802.00801 [quant-ph]} \BibitemShut
  {NoStop}%
\bibitem [{\citenamefont {{Sahu}}\ \emph {et~al.}(2018)\citenamefont {{Sahu}},
  \citenamefont {{Xu}},\ and\ \citenamefont {{Swingle}}}]{sahu2018scrambling}%
  \BibitemOpen
  \bibfield  {author} {\bibinfo {author} {\bibfnamefont {S.}~\bibnamefont
  {{Sahu}}}, \bibinfo {author} {\bibfnamefont {S.}~\bibnamefont {{Xu}}}, \ and\
  \bibinfo {author} {\bibfnamefont {B.}~\bibnamefont {{Swingle}}},\ }\href@noop
  {} {\bibfield  {journal} {\bibinfo  {journal} {ArXiv e-prints}\ } (\bibinfo
  {year} {2018})},\ \Eprint {http://arxiv.org/abs/1807.06086} {arXiv:1807.06086
  [cond-mat.str-el]} \BibitemShut {NoStop}%
\bibitem [{\citenamefont {Bravyi}\ and\ \citenamefont
  {Kitaev}(2002)}]{bravyi2002fermionic}%
  \BibitemOpen
  \bibfield  {author} {\bibinfo {author} {\bibfnamefont {S.~B.}\ \bibnamefont
  {Bravyi}}\ and\ \bibinfo {author} {\bibfnamefont {A.~Y.}\ \bibnamefont
  {Kitaev}},\ }\href {\doibase https://doi.org/10.1006/aphy.2002.6254}
  {\bibfield  {journal} {\bibinfo  {journal} {Annals of Physics}\ }\textbf
  {\bibinfo {volume} {298}},\ \bibinfo {pages} {210 } (\bibinfo {year}
  {2002})}\BibitemShut {NoStop}%
\bibitem [{\citenamefont {Chapman}\ and\ \citenamefont
  {Miyake}(2018)}]{chapman2017classical}%
  \BibitemOpen
  \bibfield  {author} {\bibinfo {author} {\bibfnamefont {A.}~\bibnamefont
  {Chapman}}\ and\ \bibinfo {author} {\bibfnamefont {A.}~\bibnamefont
  {Miyake}},\ }\href {\doibase 10.1103/PhysRevA.98.012309} {\bibfield
  {journal} {\bibinfo  {journal} {Phys. Rev. A}\ }\textbf {\bibinfo {volume}
  {98}},\ \bibinfo {pages} {012309} (\bibinfo {year} {2018})}\BibitemShut
  {NoStop}%
\bibitem [{\citenamefont {Anderson}(1958)}]{anderson1958absence}%
  \BibitemOpen
  \bibfield  {author} {\bibinfo {author} {\bibfnamefont {P.~W.}\ \bibnamefont
  {Anderson}},\ }\href {\doibase 10.1103/PhysRev.109.1492} {\bibfield
  {journal} {\bibinfo  {journal} {Phys. Rev.}\ }\textbf {\bibinfo {volume}
  {109}},\ \bibinfo {pages} {1492} (\bibinfo {year} {1958})}\BibitemShut
  {NoStop}%
\bibitem [{\citenamefont {Hosur}\ \emph {et~al.}(2016)\citenamefont {Hosur},
  \citenamefont {Qi}, \citenamefont {Roberts},\ and\ \citenamefont
  {Yoshida}}]{hosur2016chaos}%
  \BibitemOpen
  \bibfield  {author} {\bibinfo {author} {\bibfnamefont {P.}~\bibnamefont
  {Hosur}}, \bibinfo {author} {\bibfnamefont {X.-L.}\ \bibnamefont {Qi}},
  \bibinfo {author} {\bibfnamefont {D.~A.}\ \bibnamefont {Roberts}}, \ and\
  \bibinfo {author} {\bibfnamefont {B.}~\bibnamefont {Yoshida}},\ }\href
  {\doibase 10.1007/JHEP02(2016)004} {\bibfield  {journal} {\bibinfo  {journal}
  {Journal of High Energy Physics}\ }\textbf {\bibinfo {volume} {2016}},\
  \bibinfo {pages} {4} (\bibinfo {year} {2016})}\BibitemShut {NoStop}%
\bibitem [{\citenamefont {Cotler}\ \emph {et~al.}(2017)\citenamefont {Cotler},
  \citenamefont {Hunter-Jones}, \citenamefont {Liu},\ and\ \citenamefont
  {Yoshida}}]{cotler2017chaos}%
  \BibitemOpen
  \bibfield  {author} {\bibinfo {author} {\bibfnamefont {J.}~\bibnamefont
  {Cotler}}, \bibinfo {author} {\bibfnamefont {N.}~\bibnamefont
  {Hunter-Jones}}, \bibinfo {author} {\bibfnamefont {J.}~\bibnamefont {Liu}}, \
  and\ \bibinfo {author} {\bibfnamefont {B.}~\bibnamefont {Yoshida}},\ }\href
  {\doibase 10.1007/JHEP11(2017)048} {\bibfield  {journal} {\bibinfo  {journal}
  {Journal of High Energy Physics}\ }\textbf {\bibinfo {volume} {2017}},\
  \bibinfo {pages} {48} (\bibinfo {year} {2017})}\BibitemShut {NoStop}%
\bibitem [{\citenamefont {{Knill}}(2001)}]{knill2001fermionic}%
  \BibitemOpen
  \bibfield  {author} {\bibinfo {author} {\bibfnamefont {E.}~\bibnamefont
  {{Knill}}},\ }\href@noop {} {\bibfield  {journal} {\bibinfo  {journal}
  {eprint arXiv:quant-ph/0108033}\ } (\bibinfo {year} {2001})},\ \Eprint
  {http://arxiv.org/abs/quant-ph/0108033} {quant-ph/0108033} \BibitemShut
  {NoStop}%
\bibitem [{\citenamefont {Terhal}\ and\ \citenamefont
  {DiVincenzo}(2002)}]{terhal2002classical}%
  \BibitemOpen
  \bibfield  {author} {\bibinfo {author} {\bibfnamefont {B.~M.}\ \bibnamefont
  {Terhal}}\ and\ \bibinfo {author} {\bibfnamefont {D.~P.}\ \bibnamefont
  {DiVincenzo}},\ }\href {\doibase 10.1103/PhysRevA.65.032325} {\bibfield
  {journal} {\bibinfo  {journal} {Phys. Rev. A}\ }\textbf {\bibinfo {volume}
  {65}},\ \bibinfo {pages} {032325} (\bibinfo {year} {2002})}\BibitemShut
  {NoStop}%
\bibitem [{\citenamefont {Bravyi}(2006)}]{bravyi2006universal}%
  \BibitemOpen
  \bibfield  {author} {\bibinfo {author} {\bibfnamefont {S.}~\bibnamefont
  {Bravyi}},\ }\href {\doibase 10.1103/PhysRevA.73.042313} {\bibfield
  {journal} {\bibinfo  {journal} {Phys. Rev. A}\ }\textbf {\bibinfo {volume}
  {73}},\ \bibinfo {pages} {042313} (\bibinfo {year} {2006})}\BibitemShut
  {NoStop}%
\bibitem [{\citenamefont {Jozsa}\ and\ \citenamefont
  {Miyake}(2008)}]{jozsa2008matchgates}%
  \BibitemOpen
  \bibfield  {author} {\bibinfo {author} {\bibfnamefont {R.}~\bibnamefont
  {Jozsa}}\ and\ \bibinfo {author} {\bibfnamefont {A.}~\bibnamefont {Miyake}},\
  }\href {\doibase 10.1098/rspa.2008.0189} {\bibfield  {journal} {\bibinfo
  {journal} {Proceedings of the Royal Society of London A: Mathematical,
  Physical and Engineering Sciences}\ }\textbf {\bibinfo {volume} {464}},\
  \bibinfo {pages} {3089} (\bibinfo {year} {2008})}\BibitemShut {NoStop}%
\bibitem [{\citenamefont {de~Melo}\ \emph {et~al.}(2013)\citenamefont
  {de~Melo}, \citenamefont {Ćwikliński},\ and\ \citenamefont
  {Terhal}}]{melo2013power}%
  \BibitemOpen
  \bibfield  {author} {\bibinfo {author} {\bibfnamefont {F.}~\bibnamefont
  {de~Melo}}, \bibinfo {author} {\bibfnamefont {P.}~\bibnamefont
  {Ćwikliński}}, \ and\ \bibinfo {author} {\bibfnamefont {B.~M.}\
  \bibnamefont {Terhal}},\ }\href
  {http://stacks.iop.org/1367-2630/15/i=1/a=013015} {\bibfield  {journal}
  {\bibinfo  {journal} {New Journal of Physics}\ }\textbf {\bibinfo {volume}
  {15}},\ \bibinfo {pages} {013015} (\bibinfo {year} {2013})}\BibitemShut
  {NoStop}%
\bibitem [{\citenamefont {{Brod}}\ and\ \citenamefont
  {{Childs}}(2013)}]{brod2014computational}%
  \BibitemOpen
  \bibfield  {author} {\bibinfo {author} {\bibfnamefont {D.~J.}\ \bibnamefont
  {{Brod}}}\ and\ \bibinfo {author} {\bibfnamefont {A.~M.}\ \bibnamefont
  {{Childs}}},\ }\href@noop {} {\bibfield  {journal} {\bibinfo  {journal}
  {ArXiv e-prints}\ } (\bibinfo {year} {2013})},\ \Eprint
  {http://arxiv.org/abs/1308.1463} {arXiv:1308.1463 [quant-ph]} \BibitemShut
  {NoStop}%
\bibitem [{\citenamefont {Sekino}\ and\ \citenamefont
  {Susskind}(2008)}]{sekino2008fast}%
  \BibitemOpen
  \bibfield  {author} {\bibinfo {author} {\bibfnamefont {Y.}~\bibnamefont
  {Sekino}}\ and\ \bibinfo {author} {\bibfnamefont {L.}~\bibnamefont
  {Susskind}},\ }\href {http://stacks.iop.org/1126-6708/2008/i=10/a=065}
  {\bibfield  {journal} {\bibinfo  {journal} {Journal of High Energy Physics}\
  }\textbf {\bibinfo {volume} {2008}},\ \bibinfo {pages} {065} (\bibinfo {year}
  {2008})}\BibitemShut {NoStop}%
\bibitem [{\citenamefont {{Farhi}}\ \emph {et~al.}(2014)\citenamefont
  {{Farhi}}, \citenamefont {{Goldstone}},\ and\ \citenamefont
  {{Gutmann}}}]{farhi2014quantum}%
  \BibitemOpen
  \bibfield  {author} {\bibinfo {author} {\bibfnamefont {E.}~\bibnamefont
  {{Farhi}}}, \bibinfo {author} {\bibfnamefont {J.}~\bibnamefont
  {{Goldstone}}}, \ and\ \bibinfo {author} {\bibfnamefont {S.}~\bibnamefont
  {{Gutmann}}},\ }\href@noop {} {\bibfield  {journal} {\bibinfo  {journal}
  {ArXiv e-prints}\ } (\bibinfo {year} {2014})},\ \Eprint
  {http://arxiv.org/abs/1411.4028} {arXiv:1411.4028 [quant-ph]} \BibitemShut
  {NoStop}%
\end{thebibliography}
\end{document}